\documentclass[letterpaper,twocolumn,twoside]{IEEEtran} 
\usepackage{amsmath, amssymb, bm, cite, epsfig, psfrag}
\usepackage{algorithm,algorithmic}
\usepackage{bbm}
\usepackage[usenames]{color}
\usepackage{tikz}
\usepackage{etoolbox}
\usepackage{enumerate}
\usetikzlibrary{shapes,arrows}
\usetikzlibrary{patterns}
\usepackage{pgfplots}
\usepackage{url,hyperref}
\usepackage{xspace}
\usepackage{enumerate}

\newtoggle{conference}
\toggletrue{conference} 

\usepackage[normalem]{ulem} 

\newcommand{\textb}[1]{\textcolor{black}{#1}}

\def\e{\mathrm{e}}
\def\d{\mathrm{\, d}}

\def\beq{\begin{equation}}
\def\eeq{\end{equation}}
\def\beqa{\begin{eqnarray}}
\def\eeqa{\end{eqnarray}}
\def\beqan{\begin{eqnarray*}}
\def\eeqan{\end{eqnarray*}}

\def\N{{\mathbb{N}}}

\def\R{{\mathbb{R}}}

\def\argmin{\mathop{\mathrm{arg\,min}}}
\def\argmax{\mathop{\mathrm{arg\,max}}}

\def\Diag{\mathop{\mathrm{Diag}}}
\def\x{\times}

\newtheorem{theorem}{Theorem}
\newtheorem{lemma}{Lemma}

\newtheorem{assumption}{Assumption}

\def\bhat{\widehat{b}}

\def\gtilde{\tilde{g}}

\def\xhat{\widehat{x}}

\def\zhat{\widehat{z}}

\def\arr{\rightarrow}
\def\Exp{\mathbb{E}}
\def\var{\mbox{var}}

\def\tm1{t\! - \! 1}
\def\tp1{t\! + \! 1}
\def\km1{k\! - \! 1}
\def\kp1{k\! + \! 1}

\newcommand{\zero}{\mathbf{0}}
\newcommand{\one}{\mathbf{1}}
\newcommand{\abf}{\mathbf{a}}
\newcommand{\bbf}{\mathbf{b}}
\newcommand{\bbfhat}{\widehat{\mathbf{b}}}
\newcommand{\dbf}{\mathbf{d}}
\newcommand{\ebf}{\mathbf{e}}
\newcommand{\gbf}{\mathbf{g}}

\newcommand{\pbf}{\mathbf{p}}

\newcommand{\qbf}{\mathbf{q}}
\newcommand{\qbfhat}{\widehat{\mathbf{q}}}
\newcommand{\rbf}{\mathbf{r}}

\newcommand{\sbf}{\mathbf{s}}
\newcommand{\sbfbar}{\overline{\mathbf{s}}}

\newcommand{\ubf}{\mathbf{u}}

\newcommand{\vbf}{\mathbf{v}}

\newcommand{\wbf}{\mathbf{w}}

\newcommand{\xbf}{\mathbf{x}}
\newcommand{\xbfhat}{\widehat{\mathbf{x}}}

\newcommand{\ybf}{\mathbf{y}}
\newcommand{\zbf}{\mathbf{z}}
\newcommand{\zbfbar}{\overline{\mathbf{z}}}
\newcommand{\zbfhat}{\widehat{\mathbf{z}}}
\newcommand{\Abf}{\mathbf{A}}
\newcommand{\Bbf}{\mathbf{B}}

\newcommand{\Dbf}{\mathbf{D}}

\newcommand{\Hbf}{\mathbf{H}}

\newcommand{\Ibf}{\mathbf{I}}
\newcommand{\Jbf}{\mathbf{J}}

\newcommand{\Pbf}{\mathbf{P}}

\newcommand{\Sbf}{\mathbf{S}}
\newcommand{\Ubf}{\mathbf{U}}

\def\gammabf{{\boldsymbol \gamma}}

\def\xibf{{\boldsymbol \xi}}
\def\taubf{{\boldsymbol \tau}}
\def\taubar{{\overline{\tau}}}
\def\taubfbar{{\overline{\boldsymbol \tau}}}
\def\taubfhat{{\widehat{\boldsymbol \tau}}}
\def\phihat{{\widehat{\phi}}}

\newcommand{\thetabar}{{\overline{\theta}}}
\newcommand{\thetabf}{{\bm{\theta}}}

\newcommand{\SP}{MMSE\xspace}
\newcommand{\MS}{MAP\xspace}

\newcommand{\tran}{^{\text{\sf T}}}

\newcommand{\fz}{f_z}
\newcommand{\fzi}{f_{z_i}}
\newcommand{\fx}{f_x}
\newcommand{\fxj}{f_{x_j}}
\newcommand{\Zz}{Z_z}

\newcommand{\ZzT}{Z_{z,T}}
\newcommand{\ZxT}{Z_{x,T}}
\newcommand{\pdf}{p}
\newcommand{\ALG}{ADMM-GAMP\xspace}
\newcommand{\defn}{\triangleq}
\newcommand{\const}{\text{const}}

\newcommand{\Hb}{\mathcal{H}_b}

\tikzstyle{block}=[rectangle,draw, fill=blue!20,
    minimum height=1cm, minimum width=1.5cm]
\tikzstyle{signal}=[coordinate,draw]

\title{Inference for Generalized Linear Models via Alternating Directions and Bethe Free Energy Minimization}

  \author{
     Sundeep Rangan, \IEEEmembership{Fellow,~IEEE},
     Alyson K. Fletcher, \IEEEmembership{Member,~IEEE},\\
     Philip Schniter, \IEEEmembership{Fellow,~IEEE}, and
     Ulugbek S. Kamilov \IEEEmembership{Member,~IEEE}
     \thanks{S. Rangan (email: srangan@poly.edu) is with
           the Department of Electrical and Computer Engineering,
           Polytechnic Institute of New York University, Brooklyn, NY\@.
	   His work was supported by the National Science
	   Foundation under Grant No.~1116589.}
     \thanks{A.~K.~Fletcher (email: akfletcher@ucla.edu) is with
           the Departments of Statistics, Mathematics, and Electrical Engineering,
           University of California, Los Angeles.}
     \thanks{P.~Schniter (email: schniter@ece.osu.edu) is with
           the Department of Electrical and Computer Engineering,
           The Ohio State University.
	   His work was supported in part by
	   the National Science Foundation Grants CCF-1218754, CCF-1018368, and CCF-1527162.}
     \thanks{U.~S.~Kamilov (email: ulugbek.kamilov@epfl.ch) is
	   with the Biomedical Imaging Group, {\'{E}}cole
	   polytechnique f{\'{e}}d{\'{e}}rale de Lausanne
	   (EPFL), CH-1015 Lausanne VD, Switzerland. His work was
	   supported by the European Research Council under the
	   European Union's Seventh Framework Programme
	   (FP7/2007-2013)/ERC Grant Agreement 267439.}
     \thanks{This work was presented in part at the 2015 IEEE Symposium on Information Theory.}
    }

\begin{document}
\setlength{\arraycolsep}{0.5mm}

\maketitle
\begin{abstract}
Generalized Linear Models (GLMs), where a random vector $\xbf$
is observed through a noisy, possibly nonlinear,
function of a linear transform $\zbf=\Abf\xbf$, arise in a range
of applications in nonlinear filtering and regression.
Approximate Message Passing (AMP) methods, based on
loopy belief propagation, are a promising
class of approaches for approximate inference in these
models.  AMP methods are computationally simple, general,
and admit precise analyses with testable conditions for optimality
for large i.i.d.\ transforms $\Abf$.  However, the algorithms
can diverge for general $\Abf$.  This paper
presents a convergent approach to the generalized AMP (GAMP)
algorithm based on direct minimization
of a large-system limit approximation of the
Bethe Free Energy (LSL-BFE).
The proposed method uses a double-loop procedure, where the outer loop
successively linearizes the LSL-BFE and the inner loop
minimizes the linearized LSL-BFE using the Alternating Direction
Method of Multipliers (ADMM).  The proposed
method, called \ALG,
is similar in structure to the original GAMP method, but with an
additional least-squares minimization.   It is shown that
for strictly convex, smooth penalties, \ALG
is guaranteed to converge to a local minimum of the LSL-BFE,
thus providing a convergent alternative to GAMP
that is stable under arbitrary transforms.  Simulations are also presented
that demonstrate the robustness of the method for non-convex
penalties as well.
\end{abstract}


\section{Introduction} \label{sec:intro}

\begin{figure}
\center
\begin{tikzpicture}[scale=1]
    \node (x) {$\xbf \sim p_{\xbf}$};
    \node [block,node distance=2.5cm]  (A)   [right of=x] {$\Abf$ };
    \node [block,node distance=2.5cm]  (pyz) [right of=A] {$p_{\ybf|\zbf}$}
        edge [<-] node[auto,swap] {$\zbf$} (A);
    \node [node distance=1.5cm] (y) [right of=pyz] {$\ybf$};

    \node [below of=x,font=\footnotesize] 
        {\parbox{2.0cm}{\centering Unknown input,\\ independent components} };
    \node [below of=A,font=\footnotesize] {Linear transform};
    \node [below of=pyz,text width=1.7cm,font=\footnotesize,xshift=-0.15cm]
        {\parbox{1.7cm}{\centering Componentwise\\ \hspace{0.1cm} output map} };

    \draw [->] (x) -- (A);
    \draw [->] (pyz) -- (y);
\end{tikzpicture}
\caption{Generalized Linear Model (GLM) where an unknown random vector $\xbf$
is observed via a linear transform followed by componentwise likelihood to yield
a measurement vector $\ybf$.  \label{fig:linMixMod} }
\end{figure}
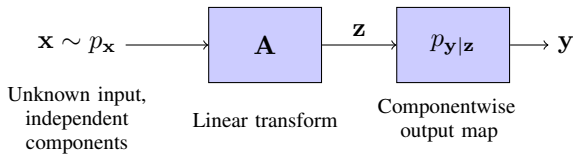

Consider the problem of estimating a random vector $\xbf \in \R^n$
from observations $\ybf \in \R^m$ as shown in Fig.~\ref{fig:linMixMod}.
The unknown vector is assumed to have
a prior density of the form $p(\xbf)=\e^{-\fx(\xbf)}$ and the
observations $\ybf \in \R^m$ are described by a likelihood function of the form $\pdf(\ybf|\xbf)=\e^{-\fz(\Abf\xbf,\ybf)}$ for some known transform $\Abf \in \R^{m \x n}$.
In statistics, this model is a special case of
a generalized linear model (GLM)
\cite{NelWed:72,McCulNel:89} and
arises in a range of applications including statistical regression, filtering, inverse problems, and nonlinear forms of compressed sensing.
The posterior density of $\xbf$ given $\ybf$ in the GLM model
is given by
\beq \label{eq:pxya}
    \pdf_{\xbf|\ybf}(\xbf|\ybf) = \frac{1}{Z(\ybf)} \exp\left[ -\fx(\xbf) - \fz(\Abf\xbf,\ybf) \right],
\eeq
where $Z(\ybf)$ is a normalization constant.
\textb{In the sequel, we will often omit the dependence on $\ybf$ and simply write
\beq \label{eq:pxy}
    \pdf_{\xbf|\ybf}(\xbf|\ybf) = \frac{1}{Z} \exp\left[ -\fx(\xbf) - \fz(\Abf\xbf) \right],
\eeq
so that the dependence on $\ybf$ in the function $\fz(\cdot)$ and the normalization constant
$Z$ is implicit.}
In this work, we consider the inference problem of estimating
the posterior marginal distributions,
$\pdf_{x_j|\ybf}(x_j|\ybf)$.
From these posterior marginals, one can compute the posterior means and variances
\begin{subequations}
\label{eq:xhatMMSE}
\beqa
\label{eq:xhatMMSE_a}
    \xhat_j    & \defn & \Exp(x_j|\ybf), \\
\label{eq:xhatMMSE_b}
    \tau_{x_j} & \defn & \var(x_j|\ybf).
\eeqa
\end{subequations}
We study this inference problem in the case where
the functions $\fx$ and $\fz$ are separable, in that they are of the form
\begin{subequations}
\label{eq:fxzsep}
\beqa
\label{eq:fxzsep_x}
    \fx(\xbf) &=& \sum_{j=1}^n \fxj(x_j), \\
\label{eq:fxzsep_z}
    \fz(\zbf) &=& \sum_{i=1}^m \fzi(z_i),
\eeqa
\end{subequations}
for some scalar functions $\fxj$ and $\fzi$.
The separability assumption \eqref{eq:fxzsep_x} corresponds to
the components in $\xbf$ being {\em a priori} independent.
\textb{Recalling the implicit dependence of $\fz$ on $\ybf$,
the separability assumption \eqref{eq:fxzsep_z} corresponds to}
the observations $\ybf$ being
conditionally independent given the transform outputs $\zbf\defn\Abf\xbf$.

For posterior densities of the form \eqref{eq:pxy},
there are several computationally efficient
methods to find the \emph{maximum a posteriori} (MAP)
estimate, which is given by
\beqa
     \xbfhat = \argmax_{\xbf}\pdf_{\xbf|\ybf}(\xbf|\ybf)
    = \argmin_\xbf \left[ \fx(\xbf) + \fz(\Abf\xbf)\right]. \label{eq:xhatMAP}
\eeqa
Under the separability assumptions \eqref{eq:fxzsep},
the MAP minimization \eqref{eq:xhatMAP} admits a factorizable dual
decomposition that can be exploited by a variety
of approaches, including variants of the
iterative shrinkage and thresholding algorithm (ISTA)
\cite{ChamDLL:98,DaubechiesDM:04,WrightNF:09,BeckTeb:09,Nesterov:07,BioDFig:07}
and the alternating direction method of multipliers (ADMM)
\cite{BoydPCPE:09,Esser:JIS:10,Chambolle:JMIV:11,He:JIS:12}.

In contrast, the inference problem of
estimating the posterior marginals $\pdf(x_j|\ybf)$
and the corresponding \textb{minimum mean squared error (MMSE)} estimates \eqref{eq:xhatMMSE_a}
is often more difficult---even in the case when $\fx$ and $\fz$
are convex.  As a simple example, consider the case where $\fx(\xbf)=0$
and
each $\fzi(z_i)$ constrains $z_i$ to belong to some interval, so that $\fz(\Abf\xbf)$
constrains $\xbf$ to belong to some polytope.  The MAP estimate
\eqref{eq:xhatMAP} is then given by any point
in the polytope.  Such a
point can can be computed via a linear program.  However, the MMSE estimate \eqref{eq:xhatMMSE_a}
is the centroid of the polytope which is, in general, \#P-hard to compute
\cite{rademacher2007approximating}.

GLM inference methods often use
a penalized quasi-likelihood method \cite{breslow1993approximate}
or some form
of Gibbs sampling \cite{zeger1991generalized,gamerman1997sampling}.
In recent years, Bayesian forms of approximate message passing (AMP)
have been considered as a potential alternate class of methods
for approximate inference in GLMs
\cite{DonohoMM:09,DonohoMM:10-ITW1,DonohoMM:10-ITW2,Rangan:10-CISS,Rangan:10arXiv-GAMP,Rangan:11-ISIT}.
AMP methods are based on Gaussian and quadratic approximations to loopy
belief propagation (loopy BP) in graphical models
and are both computationally simple and
applicable to arbitrary separable penalty functions $\fx$ and
$\fz$.  In addition, for certain large i.i.d.\ transforms
$\Abf$, they have the benefit that the behavior of the algorithm can be exactly predicted
by a state evolution analysis, which then provides
testable conditions for Bayes optimality
\cite{BayatiM:11,Rangan:11-ISIT,JavMon:12-arXiv}.

Unfortunately, for general $\Abf$, AMP methods may diverge \cite{RanSchFle:14-ISIT,Caltagirone:14-ISIT}---a situation
that is not surprising given that AMP is based on loopy BP, which also
may diverge.  Several recent modifications have been proposed
to improve the stability of AMP, including
damping \cite{RanSchFle:14-ISIT}, sequential updating \cite{manoel2014swamp},
and adaptive damping \cite{Vila:ICASSP:15}.
However, while these methods
appear to perform well empirically, little has been proven rigorously
about their convergence.

The main goal of this paper is to provide a
provably convergent approach to AMP\@.
We focus on the generalized AMP (GAMP) method of \cite{Rangan:11-ISIT},
which allows arbitrary
separable functions for both $\fx$ and $\fz$.
Our approach to stabilizing GAMP is based on reconsidering the inference
problem as a type of free-energy minimization.
Specifically, it is known that GAMP can be considered as an
iterative procedure for minimizing a large-system-limit approximation of the so-called Bethe Free Energy (BFE) \cite{RanSRFC:13-ISIT,Krzakala:14-ISITbethe}, which we abbreviate as ``LSL-BFE" in the sequel.
The BFE also plays a central role in loopy BP \cite{YedidiaFW:03}, and
we review both the BFE and LSL-BFE in Section~\ref{sec:Bethe}.

In contrast to GAMP, which \emph{implicitly} minimizes the LSL-BFE through an approximation of the sum-product algorithm, our proposed approach \emph{explicitly} minimizes the LSL-BFE\@.
We propose a double-loop algorithm, similar to
the well-known Convex Concave Procedure (CCCP)
\cite{yuille2002concave}.  Specifically,
the outer loop of our method
successively approximates the LSL-BFE by partially linearizing the LSL-BFE
around the current belief estimate, while
the inner loop minimizes the
linearized LSL-BFE using ADMM
\cite{BoydPCPE:09}.  Similar applications of ADMM
have also been proposed for related free-energy minimizations
\cite{yedidia2011alternating,ibrahimi2011robust}.
Interestingly, our proposed double-loop algorithm, which we dub
\ALG, is similar in structure to the original GAMP method
of \cite{Rangan:11-ISIT}, but with an additional least squares optimization.
We discuss these differences in detail in Section~\ref{sec:gamp}.

Our main theoretical result shows that, for strictly convex penalties,
the proposed \ALG algorithm is guaranteed to converge to at least
a local minimum of the LSL-BFE\@.  In this way, we obtain a variant of the GAMP
method with a provable convergence guarantee for arbitrary transforms $\Abf$.
In addition, using hardening arguments
similar to \cite{Tanaka:02,RanganFG:09}, we show that our \ALG
can also be applied to the MAP estimation problem, in which case we
can obtain global convergence for strictly convex, smooth penalties.
Also, while our theory requires convex penalties, we present
simulations that show robust behavior even in non-convex cases.

\color{black}
\section{The GLM and Examples} \label{sec:GLM}

Before describing our optimization approach, it is useful to briefly provide some
examples of the model~\eqref{eq:pxya} to illustrate the generality of the framework.
As a first simple example, consider a simple linear model
\beq \label{eq:yawgn}
    \ybf = \Abf\xbf + \mathbf{\epsilon}, , 
\eeq
where $\Abf$ is a known matrix, $\xbf$ is an unknown vector and $\epsilon$ is a noise vector.
In statistics, $\Abf$ would be the data matrix with predictors, $\xbf$ would be the vector of 
regression coefficients, $\ybf$ the vector of target or response variables 
and $\mathbf{\epsilon}$ would
represent the model errors.  
To place this model in the framework of this paper, we must impose a prior $\pdf(\xbf)$ on $\xbf$
and model the noise $\mathbf{\epsilon}$ as a random vector 
independent of $\Abf$ and $\xbf$.  Under these assumptions, the posterior density 
of $\xbf$ given $\ybf$ will be of the form \eqref{eq:pxya} if we define
\beq \label{eq:fxzex}
    \fx(\xbf) := -\ln \pdf(\xbf), \quad
    \fz(\zbf) := - \ln \pdf(\ybf|\zbf) = -\ln \pdf_{\epsilon}(\ybf-\zbf).
\eeq
The separability assumption \eqref{eq:fxzsep} will be valid if the components of $x_j$ are
$w_i$ are independent so the prior and noise density
factorizes as 
\[
    \pdf(\xbf) = \prod_{j=1}^n \pdf(x_j), \quad 
    \pdf(\mathbf{\epsilon}) = \prod_{i=1}^m \pdf(\epsilon_i).
\]
If the output noise $\mathbf{\epsilon}$ is Gaussian with independent
components $\epsilon_i \sim {\mathcal N}(0,\sigma^2_\epsilon)$, the output factor
$\fz(\zbf)$ in \eqref{eq:fxzex} has a quadratic cost,
\[
    \fz(\zbf) = \frac{1}{2\sigma^2_\epsilon} \|\ybf-\zbf\|^2.
\]
Similarly, if $\xbf$ has a Gaussian prior with ${\mathcal N}(0,\sigma^2_x\Ibf)$,
the input factor $\fx(\xbf)$ will be given by
\[
    \fx(\xbf) =  \frac{1}{2\sigma^2_\epsilon} \|\xbf\|^2.
\]
Note that the estimation in this quadratic case would be given by standard least
squares estimation.  

However, much more general models are possible.  For example, for Bayesian
forms of compressed sensing problems~\cite{JiXueCarin:08}, 
one can impose a sparse prior $\pdf(\xbf)$ such as
a Bernoulli-Gaussian or a heavy-tailed density.  

Also, for the output, any likelihood
$\pdf(\ybf|\zbf)$ that factorizes as $\prod_i \pdf(y_i|z_i)$ can be incorporated.
This model would occur, for example, under any output nonlinearities
as considered in \cite{blumensath2013compressed},
\[
    y_i = \phi_i(z_i) + \epsilon_i, 
\]
where $\phi_i(z_i)$ is a known, nonlinear function and $\epsilon_i$ is noise.
The  model can also include logistic regression \cite{Bishop:06} where $y_i \in \{0,1\}$ 
is a binary class variable and 
\[
    P(y_i=1|z_i) = 1-P(y_i=0|z_i) = \sigma(z_i),
\]
for some sigmoidal function $\sigma(z)$.  One-bit and quantized compressed sensing
\cite{KamilovGR:12} as well as Poisson output models \cite{KamRanFU:12-IT} can also
be easily modeled.
\color{black}

\section{Bethe Free Energy Minimization} \label{sec:Bethe}

We next provide a brief review of the Bethe Free Energy (BFE) minimization approach to
estimation of marginal densities in GLMs.
A more complete treatment of this topic, along with related ideas
in variational inference, can be found in \cite{YedidiaFW:03,WainwrightJ:08}.

For a generic density $\pdf(\xbf|\ybf)$,
exact computation of the marginal densities
$\pdf(x_j|\ybf)$ is difficult, because it involves a potentially high-dimensional integration.
BFE minimization provides an approximate
approach to marginal density computation for the case when
the joint density admits a factorizable structure of the form
\beq \label{eq:pfact}
    p(\xbf|\ybf) \propto \prod_{\ell=1}^L \psi_\ell(\xbf_{\alpha(\ell)}|\ybf),
\eeq
where, for each $\ell$, $\xbf_{\alpha(\ell)}$ is a sub-vector
of $\xbf$ created from indices in the subset $\alpha(\ell)$ and
$\psi_\ell$ is a potential function on that sub-vector.
In this case, BFE minimization aims to compute the vectors of densities
\beqan
    \bbf &\defn& [b_1,\dots,b_n]\tran
    \quad \text{and} \quad
    \qbf \defn [q_1,\dots,q_L]\tran ,
\eeqan
where $b_j(x_j)$ represents an estimate of the marginal density $p(x_j|\ybf)$
and where $q_\ell(\xbf_{\alpha(\ell)})$ represents an estimate
of the joint density $p(\xbf_{\alpha(\ell)}|\ybf)$ on the sub-vector $\xbf_{\alpha(\ell)}$.
These density estimates, often called ``beliefs,"
are computed using an optimization of the form
\beq \label{eq:JBFEGenOpt}
    \big(\bbfhat,\qbfhat\big) = \argmin_{(\bbf,\qbf) \in E} J(\bbf,\qbf) ,
\eeq
where $J(\bbf,\qbf)$ is the BFE given by
\beq \label{eq:JBFEGen}
    J(\bbf,\qbf) \defn
        \sum_{\ell=1}^L D(q_\ell\|\psi_\ell) + \sum_{j=1}^n (n_j-1)H(b_j);
\eeq
where $D(q_\ell\|\psi_\ell)$ is the KL divergence,
\beq \label{eq:KLdiv}
    D(a\|b) \defn \int a(x)\ln\left[ \frac{a(x)}{b(x)}\right] \d x ;
\eeq
where $H(b_j)$ is the
entropy or differential entropy; and where (for each $j$)
$n_j$ is the number of factors $\ell$
such that $j \in \alpha(\ell)$.
The BFE minimization \eqref{eq:JBFEGenOpt} is performed over the set $E$ of
all $(\bbf,\qbf)$ whose components satisfy a particular
``matching" condition:
for each $j \in \alpha(\ell)$,
the marginal density of $x_j$
within the belief $q_\ell(\xbf_{\alpha(\ell)})$ must agree with the belief
$b_j(x_j)$.
That is, the set $E$ contains all $(\bbf,\qbf)$ such that
\beq \label{eq:BFEGenCon}
    \int q_\ell(\xbf_{\alpha(\ell)}) \d \xbf_{\alpha(\ell) \backslash j}
        = b_j(x_j), \mbox{ for all } \ell,j,
\eeq
where the integration is over the components in the sub-vector $\xbf_{\alpha(\ell)}$
holding $x_j$ constant.
Note that $E$ imposes a set of linear constraints on the belief vectors $\bbf$ and $\qbf$.

The BFE minimization exactly recovers the true marginals
in certain cases (e.g., when the factor graph
has no cycles) and provides good estimates in many other
scenarios as well; see \cite{WainwrightJ:08} for a complete
discussion.  In addition, due to its separable
structure, the BFE can be typically minimized ``locally," by
solving a set of minimizations over the densities $\bbf$ and $\qbf$.
When the cardinalities of the subsets $\alpha(\ell)$ are small,
these local minimizations may involve much less computation
than directly calculating the marginals
of the full joint density $p(\xbf|\ybf)$.
In fact, the classic result of
\cite{YedidiaFW:03} is that loopy belief
propagation can be interpreted precisely as one type of iterative local
minimization of the BFE\@.

For the GLM in Section~\ref{sec:intro},
the separability assumption \eqref{eq:fxzsep} allows us to
write the density \eqref{eq:pxy} in the factorized
form \eqref{eq:pfact} using the $L=n+m$ potentials
\begin{subequations} \label{eq:psiGLM}
\beqa
    \psi_j(x_j)
    &=& \exp(-\fxj(x_j)), ~ j=1,\ldots,n, \\
    \psi_{n+i}(\xbf)
    &=& \exp(-\fzi(\abf_i\tran\xbf)),~i=1,\ldots,m,  \label{eq:psiGLMz}
\eeqa
\end{subequations}
where $\abf_i\tran$ is the $i$-th row of $\Abf$.
Note that, if $\Abf$ is a non-sparse matrix,
then $\fzi(\abf_i\tran\xbf)$
depends on all components in the vector $\xbf$.
In this case, the application of traditional loopy BP---as described for example in \cite{BaronSB:10}---does not generally
yield a significant computational improvement.

The GAMP algorithm from \cite{Rangan:11-ISIT} can be seen as an approximate
BFE minimization method for GLMs with possibly dense transforms $\Abf$.
Specifically, it was shown in \cite{RanSRFC:13-ISIT} that
the stationary points of GAMP coincide with
the local minima of the constrained optimization
\begin{subequations} \label{eq:FbetheOpt}
\beqa
    (\bhat_x,\bhat_z)
    &\defn& \argmin_{b_{x},b_{z}} J(b_x,b_z) \text{~such that}\\
    &&
        \Exp(\zbf|b_z) = \Abf\Exp(\xbf|b_x)
\eeqa
\end{subequations}
where $b_x$ and $b_z$ are product densities, i.e.,
\beq \label{eq:bxzProd}
    b_x(\xbf) = \prod_{j=1}^n b_{x_j}(x_j), \qquad
    b_z(\zbf) = \prod_{i=1}^m b_{z_i}(z_i),
\eeq
and the objective function $J(b_x,b_z)$ is given by
\beqa
    J(b_x,b_z)
    &\defn& D(b_x\|\e^{-\fx}) + D(b_z\|\Zz^{-1}\e^{-\fz})
    	\nonumber\\&&\mbox{}
        + H\big(\var(\xbf|b_x),\var(\zbf|b_z)\big),
        \label{eq:Fbethe} \\
    H(\taubf_x,\taubf_z)
    &\defn& \frac{1}{2}\sum_{i=1}^m \bigg[
        \frac{\tau_{z_i}}{\sum_{j=1}^n \!S_{ij} \tau_{x_j} } + \ln\bigg(\! 2\pi \sum_{j=1}^n S_{ij}\tau_{x_j} \!\!\bigg)
        \bigg], \qquad
        \label{eq:Hgauss} \\
    \taubf_x
    &\defn& (\tau_{x_1},\ldots,\tau_{x_n})\tran,
    	\quad \tau_{x_j} \defn \var(x_j|b_{x_j}),
    	\label{eq:taux}\\
    \taubf_z
    &\defn& (\tau_{z_1},\ldots,\tau_{z_m})\tran,
        \quad \tau_{z_i} \defn \var(z_i|b_{z_i}),
    	\label{eq:tauz}\\
    S_{ij} &=& {[\Sbf]}_{ij} \defn {[\Abf]}_{ij}^2 \quad \forall i,j.
\eeqa
Above, \textb{and in the sequel, we use $\Exp(\xbf|b_x)\in\R^n$ to denote the expectation of $\xbf$ under $\xbf\sim b_x$, and we use $\var(\xbf|b_x)\in\R^n_+$ to denote the vector whose $j$th entry is the variance of $x_j$ under $\xbf\sim b_x$. Note that $\var(\xbf|b_x)$ is not a full covariance matrix.  Also,}
$\Zz\defn\int_{\R^m} \e^{-\fz(\zbf)} \d \zbf$ is the scale factor that makes $\Zz^{-1} \e^{-\fz(\zbf)}$ a valid density over $\zbf\in\R^m$.
Although it is not essential for this paper, we note that $H\big(\var(\xbf|b_x),\var(\zbf|b_z)\big)$ is an upper bound on the differential entropy of $b_z$ that is tight when $b_z$ has independent Gaussian entries with variances $\taubf_z=\Sbf\taubf_x$.
It was then shown in \cite{Krzakala:14-ISITbethe} that the
objective function in \eqref{eq:Fbethe} can be interpreted as an approximation
of the BFE for the GLM from Section~\ref{sec:intro} in a
certain large-system limit, where $m,n \arr \infty$ and
$\Abf$ has i.i.d.\ entries.
We thus call the approximate BFE in \eqref{eq:Fbethe}
the \emph{large-system limit Bethe Free Energy} or LSL-BFE\@.

Similar to the case of loopy BP, it has been shown in
\cite{RanSRFC:13-ISIT,Krzakala:14-ISITbethe} that
the stationary points of \eqref{eq:FbetheOpt} are precisely the fixed points of sum-product GAMP\@.  Thus, GAMP can be interpreted as
an iterative procedure to find local minima of the LSL-BFE,
much in the same way that loopy BP can be interpreted as an iterative
way to find local minima of the BFE\@.
The trouble with GAMP, however, is that it does not always converge (see, e.g., the negative results in \cite{RanSchFle:14-ISIT,Caltagirone:14-ISIT,Vila:ICASSP:15}).
The situation is similar to the case of loopy BP\@.
Although several modifications of GAMP have been proposed with the goal of improving convergence, such as damping \cite{RanSchFle:14-ISIT}, sequential updating \cite{manoel2014swamp}, and adaptive damping \cite{Vila:ICASSP:15}, a globally convergent GAMP modification remains elusive.
%
%

\section{Minimization via Iterative Linearization}

Our approach to finding a convergent algorithm for minimizing
the constrained LSL-BFE employs a generalization of the
convex-concave procedure (CCCP) of \cite{yuille2002concave}
that we will refer to as \emph{Minimization via Iterative Linearization}.

\subsection{The Convex-Concave Procedure}
We first briefly review the CCCP\@.
Observe that, in the BFE \eqref{eq:JBFEGen},
the $D(q_\ell\|\psi_\ell)$ terms are convex in $q_\ell$ and the
$H(b_j)$ terms are concave in $b_j$.  Thus, the BFE \eqref{eq:JBFEGen} can be written
as a sum of terms
\[
    J(\bbf,\qbf) = f(\qbf) + h(\bbf),
\]
where $f$ is convex and $h$ is concave.
The CCCP finds a sequence of estimates of a BFE minimizer $(\bbfhat,\qbfhat)$
by iteratively linearizing the concave part of this function, i.e.,
\begin{subequations} \label{eq:cccp}
\beqa
    (\bbf^k,\qbf^k) &=& \argmin_{(\bbf,\qbf) \in E} f(\bbf) + (\gammabf^k)\tran \qbf,
    \label{eq:cccpMin} \\
    \gammabf^{\kp1} &=& \frac{\partial h(\qbf^k)}{\partial \qbf} ,
\eeqa
\end{subequations}
where $\partial h(\qbf^k)/\partial \qbf$ denotes the gradient of $h$ at $\qbf^k$.
The resulting procedure is often called a ``double-loop" algorithm,
since each iteration involves a minimization \eqref{eq:cccpMin}
that is itself usually performed by an iterative procedure.
Because $f$ is convex and the constraint $(\bbf,\qbf) \in E$ is linear,
the minimization problem \eqref{eq:cccpMin} is convex.
Thus, the CCCP converts the non-convex BFE minimization to
a sequence of convex minimizations.
In fact, it can be shown that the CCCP will
monotonically decrease the BFE for arbitrary convex $f$ and concave $h$
\cite{yuille2002concave}.

\subsection{Minimization via Iterative Linearization}
For the LSL-BFE, it is not convenient to decompose the objective
function into a convex term plus a concave term.
\textb{To handle problems like LSL-BFE minimization, we consider optimization problems of the form
}
\beq \label{eq:JoptGen}
     \min_{\bbf \in B} J(\bbf), \quad J(\bbf) = f(\bbf) + h(\gbf(\bbf)),
\eeq
where
\textb{now $\bbf$ is a vector in a Hilbert space $\Hb$,
$B$ is a closed affine subspace of $\Hb$,
$f:\Hb\rightarrow\R$ is a convex functional,
$\gbf:\Hb\rightarrow\R^p$ is a mapping from $\Hb$ to $\R^p$ for some $p\in\N$, and
$h:\R^p\rightarrow\R$ is an arbitrary functional.
Below, we use $\taubf\in\R^p$ to denote the input to $h$.
Note that}
the functionals $h$ and $h(\gbf(\cdot))$ may be neither 
concave nor convex.

\textb{To solve \eqref{eq:JoptGen},} we propose the iterative procedure shown in Algorithm~\ref{algo:IterLin}, which is reminiscent of the CCCP\@.
At each iteration $k$, an estimate $\bbf^k$ of $\argmin_{\bbf\in B}J(\bbf)$
is computed by minimizing the functional
\beq \label{eq:JlinGen}
    J(\bbf,\gammabf^k) \defn f(\bbf) + (\gammabf^k)\tran \gbf(\bbf),
\eeq
where $\gammabf^k\in\R^p$ is a ``damped" version of the gradient $\partial h(\gbf(\bbf))/\partial \bbf$.
In particular, when the \emph{damping parameter} $\theta^k$ is set
to unity, the linearization vector
is exactly equal to the gradient at $\bbf^k$, i.e., $\gammabf^k = \partial [h(\gbf(\bbf^k))]/\partial \bbf$, similar to CCCP\@.
However, in Algorithm~\ref{algo:IterLin}, we have the option of
setting $\theta^k < 1$, which has the effect of
slowing the update on $\gammabf^k$.
We will see that, by setting $\theta^k < 1$, we can
guarantee convergence when $h$ and/or $h(\gbf(\cdot))$ is non-concave.

\begin{algorithm}
\caption{Minimization via Iterative Linearization}
\begin{algorithmic}[1]  \label{algo:IterLin}
\REQUIRE{Optimization problem \eqref{eq:JoptGen}.}

\STATE{ $k \gets 0$  }
\STATE{ Select initial linearization $\gammabf^0$ }
\REPEAT
    \STATE{ } \COMMENT{Minimize the linearized function}
    \STATE{ $\bbf^k \gets \argmin_{\bbf\in B} J(\bbf,\gammabf^k)$ }
    \STATE{ }

    \STATE{ } \COMMENT{Update the linearization}
    \STATE{Select a damping parameter $\theta^k \in (0,1]$}
    \STATE{$\taubf^k \gets \gbf(\bbf^k)$}
    \STATE{$\gammabf^{\kp1} \gets (1-\theta^k)\gammabf^k + \theta^k
        \displaystyle \frac{\partial h(\taubf^k)}{\partial \taubf}$}
\UNTIL{Terminated}
\end{algorithmic}
\end{algorithm}

\subsection{Convergence of Minimization via Iterative Linearization}

Observe that when $f$ is convex, $h$ is concave, \textb{$\Hb=\R^p$ (as when $x_j$ are discrete variables)},
$\gbf$ is the identity map (i.e., $\gbf(\bbf)=\bbf$),
and there is no damping (i.e., $\theta^k=1~\forall k$),
Algorithm~\ref{algo:IterLin} reduces to the CCCP\@.
However, we are interested in possibly non-concave $h(\gbf(\cdot))$, in which case we cannot directly apply the results of \cite{yuille2002concave}.
We instead consider the following alternate conditions.

\medskip
\begin{assumption}  \label{as:IterLinConv}
Consider the optimization problem \eqref{eq:JoptGen},
and suppose that the functions $f$, $\gbf$, and $h$ have components that
are twice differentiable with uniformly bounded second derivatives.
Also, assume that there exists a convex set $\Gamma$ such that, for all
$\gammabf \in \Gamma$:
\begin{enumerate}[(a)]

\item The minimization of the linearized function,
\beq \label{eq:bhatGen}
    \bbfhat(\gammabf) \defn \argmin_{\bbf \in B} J(\bbf,\gammabf)
\eeq
exists and is unique.

\item At each minimum, the linearized objective
is uniformly strictly convex in the linear space $B$
in that there exists
constants $c_1, c_2$ with $c_2 > c_1 > 0$ such that
\beq \label{eq:Dbnd}
    c_1\|\ubf\|^2  \leq \ubf\tran\Hbf(\gammabf)\ubf \leq c_2\|\ubf\|^2,
    ~\forall \ubf: \bbfhat(\gammabf)+\ubf\in B,
\eeq
where $\Hbf(\gammabf)$ is the Hessian of $J$ with respect to $\bbf$ at $\bbfhat(\gammabf)$, i.e.,
\beq \label{eq:DdefGen}
    \Hbf(\gammabf) \defn \left.
    \frac{\partial^2 J(\bbf,\gammabf)}{\partial \bbf\, \partial \bbf\tran}
    \right|_{\bbf=\bbfhat(\gammabf)},
\eeq
and where the constants $c_1$ and $c_2$ do not depend on $\gammabf$.

\item The gradient obeys $\partial h(\gbf(\bbf)) / \partial \taubf \in \Gamma$
when $\bbf=\bbfhat(\gammabf)$.

\end{enumerate}
\end{assumption}

\medskip
\begin{theorem} \label{thm:iterLinConv}
Consider
Algorithm~\ref{algo:IterLin} under Assumption~\ref{as:IterLinConv}.
There exists a $\thetabar \in (0,1)$ such that if
the damping parameters are selected with $0 < \theta^k \leq \thetabar$ for
all $k$, and if the initialization obeys $\gammabf^0 \in \Gamma$,
then $\gammabf^k \in \Gamma$ for all $k$ and the objective monotonically
decreases, i.e.,
\beq \label{eq:JgenDec}
    J(\bbf^{\kp1}) \leq J(\bbf^k)~\forall k.
\eeq
\end{theorem}
\begin{IEEEproof} See Appendix~\ref{sec:iterConvPf}.
\end{IEEEproof}

\color{black}
The most simple case where Assumption~\ref{as:IterLinConv} holds is the setting
where $f(\bbf)$ is strictly convex and smooth, $g(\bbf)$ is linear and $h(\taubf)$ is smooth
(but neither necessarily convex nor concave).  Under these assumptions, $J(\bbf,\gammabf)$
would be strictly convex for all $\gammabf$, thereby satisfying Assumptions (a) and (b).
The assumption would also be valid for strictly convex $f(\bbf)$ and convex $g(\bbf)$,
provided we restrict to positive $\gammabf$.  In this case, to satisfy assumption (c),
we would require that $\partial h(\gbf(\bbf)) / \partial \taubf \geq 0$, i.e.\ $h(\taubf)$
is increasing in each of its component.
Interestingly, in the setting we will use below, $f(\bbf)$ will be convex, but $g(\bbf)$ will
be concave.  Nevertheless, we will show that the assumption will be satisfied.
\color{black}

\subsection{Application to LSL-BFE Minimization} \label{sec:milBFE}
To apply Algorithm~\ref{algo:IterLin} to the LSL-BFE
minimization \eqref{eq:FbetheOpt}, we first take
$B$ to be the vector of separable density pairs
$\bbf=(b_x;b_z)$
satisfying the moment matching constraint
\beq \label{eq:BdefLin}
    B = \left\{ (b_x;b_z) \mid \Exp(\zbf|b_z) = \Abf\Exp(\xbf|b_x)
    \right\}.
\eeq
Then, if we define the functions
\begin{subequations} \label{eq:fnBFE}
\beqa
    f(\bbf) \defn f(b_x,b_z)
    &=& D(b_x\|\e^{-f_x}) + D(b_z\|\Zz^{-1}\e^{-f_z}) \\
    \gbf(\bbf)\tran
    &\defn& [\var(\xbf|b_x); \var(\zbf|b_z)]
    = [\taubf_x;\taubf_z]
        \label{eq:gBFE} \qquad \\
    h([\taubf_x;\taubf_z]) &\defn& H(\taubf_x,\taubf_z), \hspace{2cm}
        \label{eq:hBFE}
\eeqa
\end{subequations}
we see that $J(b_x,b_z)$ from \eqref{eq:Fbethe}
can be cast into the form in \eqref{eq:JoptGen}.
Observe that, while $f$ is convex, the function $h(\gbf(\cdot))$
is, in general, neither convex nor concave.  Thus,
\textb{while the CCCP does not apply,} we can apply
the iterative linearization method from Algorithm~\ref{algo:IterLin}.

We will partition the linearization vector $\gammabf$ conformally
with function $\gbf$ in \eqref{eq:gBFE} as
\beq \label{eq:gamBFE}
    \gammabf = [\one./(2\taubf_r); \one./(2\taubf_p)],
\eeq
\textb{where we use ``$./$" to denote componentwise division of two vectors
and ``$;$" to denote vertical concatenation.}
The notation in \eqref{eq:gamBFE} will help to clarify the connections
to the original GAMP algorithm.
Using the above notation, the linearized objective \eqref{eq:JlinGen}
can be written as
\beqa \label{eq:JlinBFE}
    J(b_x,b_z,\taubf_r,\taubf_p)
        &\defn& D(b_x\|\e^{-f_x}) + D(b_z\|\Zz^{-1}\e^{-f_z}) \nonumber\\
        &&+ \left(\one./(2\taubf_r)\right)\tran \var(\xbf|b_x)
            \nonumber\\
        &&+ \left(\one./(2\taubf_p)\right)\tran \var(\zbf|b_z).
\eeqa

Finally, we compute the gradient $h'=\frac{\partial h}{\partial \taubf}$
of the function $h$ from
\eqref{eq:hBFE}.  Similar to $\gammabf$, we will partition the gradient
into two terms,
\beq \label{eq:taubardef}
    \one./(2\taubfbar_r) \defn
        \frac{\partial H(\taubf_x,\taubf_z)}{\partial \taubf_x},
        \quad
    \one./(2\taubfbar_p) \defn
        \frac{\partial H(\taubf_x,\taubf_z)}{\partial \taubf_z}.
\eeq
From \eqref{eq:Hgauss}, the derivative of $H$ with respect to $\tau_{z_i}$ is
\beqa
    \frac{1}{2\taubar_{p_i}} = \frac{\partial H(\taubf_x,\taubf_z)}{\partial \tau_{z_i}}
    &=& \frac{1}{2\sum_{j=1}^n S_{ij}\tau_{x_j}} .
    	\label{eq:taubarp}
\eeqa
Similarly, using the chain rule and \eqref{eq:taubarp}, we find
\beqa
    \frac{1}{2\taubar_{r_j}}
    &=& \frac{\partial H(\taubf_x,\taubf_z)}{\partial \tau_{x_j}}
    = \sum_i \frac{\partial H(\taubf_x,\taubf_z)}{\partial (\sum_k S_{ik}\tau_{x_k})}
             \frac{\partial (\sum_k S_{ik}\tau_{x_k})}{\partial \tau_{x_j}}
    \nonumber \\
    &=& \sum_{i=1}^m \frac{1}{2}\Big(-\frac{\tau_{z_i}}{\taubar_{p_i}^2}+\frac{2\pi}{2\pi \taubar_{p_i}}\Big) S_{ij} \nonumber\\
    &=& \frac{1}{2}\sum_{i=1}^m S_{ij}
    	\Big(1-\frac{\tau_{z_i}}{\taubar_{p_i}}\Big)
	\frac{1}{\taubar_{p_i}}.
    	\qquad \label{eq:taubarr}
\eeqa
We can then write \eqref{eq:taubarp} and \eqref{eq:taubarr} in vector form as
\beq
    \taubfbar_p
    = \Sbf\taubf_x,
    \quad
    \one./\taubfbar_r
    = \Sbf\tran\left[
        \left(\one-\taubf_z./\taubfbar_p\right)./\taubfbar_p
        \right]
    	\label{eq:taubarpr} .
\eeq

Substituting the above computations into the iterative linearization
algorithm, Algorithm~\ref{algo:IterLin}, we obtain
Algorithm~\ref{algo:outer}.  We refer to this as the \emph{outer loop},
since each iteration involves a minimization of the
linearized LSL-BFE in line~\ref{line:lslMin}.
We discuss this latter minimization next
and show that it can itself be performed iteratively
using a set of iterations that we will refer to as the \emph{inner loop}.

We will also show shortly that, under certain convexity conditions,
the conditions of Assumption~\ref{as:IterLinConv} are satisfied,
so that Algorithm~\ref{algo:outer} will converge to a
local minimum of the LSL-BFE\@.

\begin{algorithm}
\caption{Minimizing LSL-BFE via Iterative Linearization}
\begin{algorithmic}[1]  \label{algo:outer}
\REQUIRE{LSL-BFE objective function \eqref{eq:Fbethe}
with a matrix $\Abf$.}

\STATE{ $k \gets 0$  }
\STATE{ Select initial linearization $\taubf_p^0$, $\taubf_r^0$. }
\REPEAT
    \STATE{ } \COMMENT{Minimize the linearized LSL-BFE}
    \STATE{ $(b^k_x,b_z^k) \gets \argmin_{(b_x,b_z) \in B}
        J(b_x,b_z,\taubf_r^k,\taubf_p^k)$ } \label{line:lslMin}
    \STATE{ }

    \STATE{ } \COMMENT{Compute the gradient terms}
    \STATE{ $\taubf_x^k \gets \var(\xbf|b_x^k)$,
            $\taubf_z^k \gets \var(\zbf|b_z^k)$ }
    \STATE{ $\taubfbar_p^k \gets \Sbf\taubf_x^k$ }
    \STATE{ $\taubf_s^k \gets
        (\one - \taubf_z^k./\taubfbar_p^k)./\taubfbar_p^k$ }
    \STATE{ $\taubfbar_r^k \gets \one./(\Sbf\tran\taubf_s^k)$ }
    \STATE{ }

    \STATE{ } \COMMENT{Update the linearization}
    \STATE{Select a damping parameter $\theta^k \in (0,1]$}
    \STATE{$\one./\taubf_r^{\kp1} \gets \theta^k \one./\taubfbar_r^k
        + (1-\theta^k) \one./\taubf_r^k$}
    \STATE{$\one./\taubf_p^{\kp1} \gets \theta^k \one./\taubfbar_p^k
        + (1-\theta^k) \one./\taubf_p^k$}
\UNTIL{Terminated}
\end{algorithmic}
\end{algorithm}

\color{black}
\subsection{Alternative Methods}

While the method proposed in this paper is based on CCCP of \cite{yuille2002concave},
there are other methods for direct minimization of the BFE that may apply to the LSL-BFE as well.
For example, for problems with binary variables and pairwise penalty functions,
\cite{welling2001belief,shin2014complexity} propose a clever
re-parametrization to convert the constrained BFE minimization to an unconstrained optimization
on which gradient descent can be used.  Unfortunately, it is not obvious if the LSL-BFE
here can admit such a re-parametrization since the penalty functions are not pairwise and the
variables are not binary.
\color{black}

\section{Inner-Loop Minimization and \ALG} \label{sec:mmse}

\subsection{ADMM Principle} \label{sec:admm}

The outer loop algorithm, Algorithm~\ref{algo:outer}, requires that
in each iteration we solve a constrained optimization of the form
\beq \label{eq:FbetheLinOpt}
    (b_x,b_z) = \argmin_{b_x,b_z} J(b_x,b_z,\taubf_r,\taubf_p)
    \mbox{ s.t. } \Exp(\zbf|b_z) = \Abf\Exp(\xbf|b_x).
\eeq
We will show that this optimization can be performed by
the Alternating Direction Method of Multipliers (ADMM)
\cite{BoydPCPE:09}.
ADMM is a general approach to constrained optimizations of the form
\beq \label{eq:optGen}
    \min_{\wbf} f(\wbf) \mbox{ s.t. } \Bbf\wbf = \zero,
\eeq
where $f(\wbf)$ is an objective function and $\Bbf$ is some constraint matrix.
Corresponding to this optimization, let us define the augmented Lagrangian
\beq \label{eq:LaugGen}
    L(\wbf,\ubf;\taubf) \defn f(\wbf) + \ubf\tran\Bbf\wbf + \frac{1}{2} \|\Bbf\wbf\|^2_{\taubf},
\eeq
where $\ubf$ is a dual vector, $\taubf $ is a vector of positive weights and
$\|\xbf\|_{\taubf}^2\defn\sum_j x_j^2/\tau_j$.
The ADMM procedure then produces a sequence of estimates for the optimization
\eqref{eq:optGen} through the iterations
\begin{subequations} \label{eq:admmGen}
\beqa
    \wbf^{\tp1} &=& \argmin_{\wbf} L(\wbf,\ubf^t;\taubf) \label{eq:admmPrimal} \\
    \ubf^{\tp1} &=& \ubf^t + \Diag(\one./\taubf)\Bbf\wbf^{\tp1} \label{eq:admmDual} ,
\eeqa
\end{subequations}
\textb{where $\Diag(\dbf)$ creates a diagonal matrix from the vector $\dbf$.}
The algorithm thus alternately updates the primal variables $\wbf^t$
and dual variables $\ubf^t$. The vector $\taubf$
can be interpreted as a step-size on the primal
problem and an inverse step-size on the dual problem.

The key benefit of the ADMM method is that,
for any positive step-size vector $\taubf$,
the procedure is guaranteed to converge to a global optimum
for convex functions $f(\wbf)$ under mild conditions on $\Bbf$.

\subsection{Application of ADMM to LSL-BFE Optimization} \label{sec:admmBFE1}

The ADMM procedure can be applied to the linearized
LSL-BFE optimization \eqref{eq:FbetheLinOpt} as
follows.
First, we replace the constraint $\Exp(\zbf|b_z)=\Abf\Exp(\xbf|b_x)$ with two constraints: $\Exp(\zbf|b_z)=\Abf\vbf$ and $\Exp(\xbf|b_x)=\vbf$.
Variable splittings of this form are commonly used in the context of monotropic programming \cite{Rockafellar:Chap:81}.
With this splitting, the augmented Lagrangian for the LSL-BFE \eqref{eq:FbetheOpt} becomes
\beqa
    \lefteqn{ L(b_x,b_z,\sbf,\qbf,\vbf;\taubf_p,\taubf_r) }
    \nonumber\\
    &\defn& J(b_x,b_z,\taubf_r,\taubf_p)
    + \qbf\tran(\Exp(\xbf|b_x)-\vbf)
    + \sbf\tran(\Exp(\zbf|b_z)-\Abf \vbf)
    \nonumber \\ && \mbox{}
    + \frac{1}{2}\|\Exp(\xbf|b_x)-\vbf\|^2_{\taubf_r}
    + \frac{1}{2}\|\Exp(\zbf|b_z)-\Abf\vbf\|^2_{\taubf_p},
    \label{eq:LaugSP0}
\eeqa
where $\sbf$ and $\qbf$ represent the dual variables.
Note that the vectors
$\taubf_r$ and $\taubf_p$ that appear in the linearized LSL-BFE
$J(b_x,b_z,\taubf_r,\taubf_p)$
have been used for the augmentation terms (i.e., the last two terms)
in \eqref{eq:LaugSP0}.
This choice will be critical.
From \eqref{eq:admmGen}, the resulting ADMM recursion becomes
\begin{subequations} \label{eq:ADMM}
\beqa
    (b_x^{\tp1},b_z^{\tp1})
    &=& \argmin_{b_x,b_z} L(b_x,b_z,\sbf^t,\qbf^t,\vbf^t;\taubf_p,\taubf_r),
	\label{eq:ADMMb}\\
    \sbf^{\tp1}
    &=& \sbf^{t} + \Diag(\one./\taubf_p)\big(\Exp(\zbf|b_z^{\tp1})-\Abf\vbf^{\tp1}\big),
    	\label{eq:ADMMs}\\
    \qbf^{\tp1}
    &=& \qbf^{t} + \Diag(\one./\taubf_r)\big(\Exp(\xbf|b_x^{\tp1})-\vbf^{\tp1}\big),
    	\label{eq:ADMMq}\\
    \vbf^{\tp1}
    &=& \argmin_{\vbf} L(b_x^{\tp1},b_z^{\tp1},\sbf^{\tp1},\qbf^{\tp1},\vbf;\taubf_p,\taubf_r) . \qquad
	\label{eq:ADMMv}
\eeqa
\end{subequations}

To compute the minimization in \eqref{eq:ADMMb}, we
first note that the second and fourth terms in
\eqref{eq:LaugSP0} can be rewritten as
\beqa
    \lefteqn{ \qbf\tran(\Exp(\xbf|b_x)-\vbf)
    	+ \frac{1}{2}\|\Exp(\xbf|b_x)-\vbf\|^2_{\taubf_r}  }\nonumber\\
    &=& \sum_{j=1}^n q_j \Exp(x_j|b_{x})
	+ \frac{\Exp^2(x_j|b_{x})- 2v_j \Exp(x_j|b_x) }{2\tau_{r_j}}
    	+ \const \nonumber \\
    &\stackrel{(a)}{=}& \sum_{j=1}^n \Exp \bigg( q_j x_j
	+ \frac{x_j^2 - 2v_j x_j}{2\tau_{r_j}} \bigg| b_x \bigg)
	- \frac{ \tau_{x_j} }{2\tau_{r_j}}
    	+ \const \nonumber \\
    &=& \sum_{j=1}^n \frac{
        \Exp\big( (x_j - [v_j - \tau_{r_j}q_j])^2 \big| b_x \big)
	}{2\tau_{r_j}}
	- \frac{ \tau_{x_j}}{2\tau_{r_j}}
    	+ \const \nonumber \\
    &\stackrel{(b)}{=}& \frac{1}{2}
    	\Exp \big( \|\xbf-(\vbf-\taubf_r.\qbf)\|_{\taubf_r}^2 \big| b_x \big)
	- \sum_{j=1}^n \frac{ \tau_{x_j} }{2\tau_{r_j}}
    	+ \const , \label{eq:qprod}
\eeqa
where in (a) we used $\Exp^2(x_j|b_x)=\Exp(x_j^2|b_x)-\var(x_j|b_x)=\Exp(x_j^2|b_x)-\tau_{x_j}$;
in (b) we used ``$.$" to denote componentwise multiplication between vectors;
and ``$\const$" includes terms that are constant with respect to $b_x$ and $b_z$.
A similar development yields
\beqa
    \lefteqn{ \sbf\tran(\Exp(\zbf|b_z)-\Abf\vbf)
    	+ \frac{1}{2}\|\Exp(\zbf|b_z)-\Abf\vbf\|^2_{\taubf_p}  }\label{eq:saug}\\
    &=& \frac{1}{2}
    	\Exp \big( \|\zbf-(\Abf\vbf-\taubf_p\sbf)\|_{\taubf_p}^2 \big| b_z \big)
	- \sum_{i=1}^m \frac{ \tau_{z_i} }{2\tau_{p_i}}
    	+ \const . \nonumber
\eeqa
Also, note that the last two terms in \eqref{eq:JlinBFE} can be rewritten as
\begin{subequations} \label{eq:tauvarprod}
\beqa
    && \left(\one./(2\taubf_r)\right)\tran \var(\xbf|b_x) =
    \sum_{j=1}^n \frac{\tau_{x_j}}{2\tau_{r_j}}, \\
    && \left(\one./(2\taubf_p)\right)\tran \var(\zbf|b_z) =
    \sum_{i=1}^m \frac{\tau_{z_i}}{2\tau_{p_i}}.
\eeqa
\end{subequations}
Substituting \eqref{eq:JlinBFE},
\eqref{eq:qprod}, \eqref{eq:saug}, and \eqref{eq:tauvarprod}
into \eqref{eq:LaugSP0}, and canceling terms, we get
\beqa
    \lefteqn{
    L(b_x,b_z,\sbf,\qbf,\vbf;\taubf_p,\taubf_r)
    	}\nonumber\\
    &=& D(b_x\|\e^{-\fx})
    	+ \frac{1}{2}
    	\Exp \big( \|\xbf-(\vbf-\taubf_r.\qbf)\|_{\taubf_r}^2 \big| b_x \big)
	\nonumber\\&&\mbox{}
        + D(b_z\|\Zz^{-1}\e^{-\fz})
    	+ \frac{1}{2}
    	\Exp \big( \|\zbf-(\Abf\vbf-\taubf_p.\sbf)\|_{\taubf_p}^2 \big| b_z \big)
     \nonumber\\&&\mbox{}
        + \const \nonumber \\
    &=& \int_{\R^n} b_x(\xbf) \ln \tfrac{b_x(\xbf)}{
         \exp(
 	-\fx(\xbf) - \frac{1}{2} \|\xbf-(\vbf-\taubf_r.\qbf)\|_{\taubf_r}^2
 	) } d\xbf
 	\nonumber\\&&\mbox{}
         + \int_{\R^m} b_z(\zbf) \ln \tfrac{b_z(\zbf)}{
         \exp(
 	-\fz(\zbf) - \frac{1}{2} \|\zbf-(\Abf\vbf-\taubf_p.\sbf)\|_{\taubf_p}^2
 	) } d\zbf	
     \nonumber\\&&\mbox{}
    	+ \const \\
    &=& D(b_x\|p_x) + D(b_z\|p_z) + \const,
	\label{eq:LaugSP}
\eeqa
for $p_x(\xbf)\propto \exp(
-\fx(\xbf) - \frac{1}{2} \|\xbf-(\vbf-\taubf_r.\qbf)\|_{\taubf_r}^2)$
and $p_z(\zbf)\propto \exp(
-\fz(\zbf) - \frac{1}{2} \|\zbf-(\Abf\vbf-\taubf_p.\sbf)\|_{\taubf_p}^2)$.
Therefore, the ADMM step \eqref{eq:ADMMb} has the solution,
\begin{subequations} \label{eq:bxz}
\beqa
    b_x^{\tp1}(\xbf)
    &\propto& \exp\big( -\fx(\xbf)
        - \tfrac{1}{2} \|\xbf-\rbf^t\|_{\taubf_r}^2 \big)
	\label{eq:bx}\\
    b_z^{\tp1}(\zbf)
    &\propto& \exp\big( -\fz(\zbf)
	- \tfrac{1}{2} \|\zbf-\pbf^t\|_{\taubf_p}^2 \big) .
	\label{eq:bz}
\eeqa
\end{subequations}
for vectors
\begin{subequations} \label{eq:rpt}
\beqa
    \rbf^t
    &\defn& \vbf^t-\taubf_r.\qbf^t \\
    \pbf^t
    &\defn& \Abf\vbf^t-\taubf_p.\sbf^t ,
\eeqa
\end{subequations}
\textb{where we use ``$.$" to denote componentwise vector multiplication.}
Using Bayes rule, \eqref{eq:bx} can be interpreted as the posterior density of the random vector $\xbf$ under the prior $\e^{-\fx(\xbf)}$ and an independent Gaussian likelihood with mean $\rbf^{t}$ and variance $\taubf_r$.
Similarly, \eqref{eq:bz} can be interpreted as the posterior pdf the random vector $\zbf$ under the likelihood $\e^{-\fz(\zbf)}$ and an independent Gaussian prior with mean $\pbf^t$ and variance $\taubf_p$.

To tackle the minimization \eqref{eq:ADMMv}, we ignore the $\vbf$-invariant components in the original augmented Lagrangian \eqref{eq:LaugSP0}, after which \eqref{eq:ADMMv} can be reformulated as the least-squares problem
\beqa
    \vbf^{\tp1}
    &=& \argmin_{\vbf}
    	\|\zbf^{\tp1}+\taubf_p \sbf^{\tp1} -\Abf\vbf\|^2_{\taubf_p}
        \nonumber\\&&\mbox{}
    	+ \|\xbf^{\tp1}+\taubf_r \qbf^{\tp1} -\vbf\|^2_{\taubf_r}
    \label{eq:vmin}
\eeqa
using the definitions
\beq
    \zbf^{\tp1} \defn \Exp(\zbf|b_z^{\tp1}), \quad
    \xbf^{\tp1} \defn \Exp(\xbf|b_x^{\tp1}).
\eeq

\subsection{The \ALG Algorithm} \label{sec:ALG}

Inserting the above ADMM updates into the outer loop algorithm,
Algorithm~\ref{algo:outer}, we obtain the so-called
\ALG method summarized in Algorithm~\ref{algo:ALG}.
\textb{There and elsewhere, we use ``$.$" to denote componentwise vector-vector multiplication and ``$./$" to denote componentwise vector-vector division.}
Note that the updates for the ADMM iteration appear under the comment
``ADMM inner iteration."

Although, in principle, we should perform an infinite number
of inner-loop iterations for each outer-loop
iteration, Algorithm~\ref{algo:outer} is written in a more general
``parallel form."
In each (global) iteration $t$, there is one ADMM
update as well as one linearization update.
However, by setting the outer-loop damping parameter as $\theta^t=0$,
it is possible to bypass the linearization update.
Thus, we can obtain the desired double-loop behavior as follows:
First, hold $\theta^t=0$ for a large number of iterations,
thus running ADMM to convergence.
Then, set $\theta^t > 0$ for a single iteration
to update the linearization.  Then, hold $\theta^t=0$
for another large number of iterations, and so on.
However, the parallel form of Algorithm~\ref{algo:ALG} also
facilitates other update schedules.  For example,
we could run a small number of ADMM updates for each linearization update,
or we could run only one ADMM update per linearization update.

\color{black}
An interesting question is whether the algorithm can be run with a constant
step-size $\theta^t=\theta$ for some small $\theta$.  Unfortunately, our theoretical
analysis and numerical experiments consider only the double-loop implementation where
several ADMM iterations are run for each outer loop update.
\color{black}

\begin{algorithm}[t]
\caption{\ALG}
\begin{algorithmic}[1]  \label{algo:ALG}
\REQUIRE{
Matrix $\Abf$, estimation functions $g_x$ and $g_z$. }

\STATE{ $\Sbf \gets \Abf.\Abf$ (componentwise square)} \label{line:Sdef}
\STATE{ Initialize $\taubf_r^0 >\zero, \taubf_p^0>\zero$, $\vbf^0$ }
\STATE{ $\qbf^0 \gets \zero$, $\sbf^0 \gets \zero$ }
\STATE{ $t \gets 0$  }
\REPEAT

    \STATE{ } \COMMENT{ADMM inner iteration}
    \STATE{ $\rbf^{t} \gets \vbf^{t}-\taubf_r^{t}.\qbf^{t}$ } \label{line:rt}
    \STATE{ $\pbf^{t} \gets  \Abf\vbf^{t}-\taubf_p^{t}.\sbf^{t}$ } \label{line:pt}
    \STATE{ $\xbf^{\tp1} \gets g_x(\rbf^t,\taubf_r^t)$,~~
            $\zbf^{\tp1} \gets g_z(\pbf^t,\taubf_p^t)$} \label{line:xzest}
    \STATE{ $\qbf^{\tp1} \gets  \qbf^t+\Diag(1./\taubf_r^t)(\xbf^{\tp1}-\vbf^t)$ }
        \label{line:qt}
    \STATE{ $\sbf^{\tp1} \gets  \sbf^t+\Diag(1./\taubf_p^t)(\zbf^{\tp1}-\Abf\vbf^t)$ } \label{line:st}
    \STATE{ Compute $\vbf^{\tp1}$ from \eqref{eq:vmin} } \label{line:vt}

    \STATE{ }

    \STATE{ } \COMMENT{Compute the gradient terms}
    \STATE{ $\taubf_x^{\tp1} \gets \taubf_r^t .
             g_x'(\rbf^t,\taubf_r^t)$, ~
            $\taubf_z^{\tp1} \gets \taubf_p^t .
             g_z'(\pbf^t,\taubf_p^t)$
             } \label{line:tauxzt}
    \STATE{ $\taubfbar_p^{\tp1} \gets \Sbf\taubf_x^{\tp1}$
        \label{line:taupbart} }
    \STATE{ $\taubf_s^{\tp1} \gets
        (1 - \taubf_z^{\tp1}./\taubfbar_p^{\tp1})./\taubfbar_p^{\tp1}$
        \label{line:taust}}
    \STATE{ $\taubfbar_r^{\tp1} \gets \one./(\Sbf\tran\taubf_s^{\tp1})$
        \label{line:taurbart}}
    \STATE{ }

    \STATE{ } \COMMENT{Update the linearization}
    \STATE{Select a damping parameter $\theta^t \in [0,1]$}
    \STATE{$\one./\taubf_r^{\tp1} \gets \theta^t\one./\taubfbar_r^{\tp1}
        + (1-\theta^t)\one./\taubf_r^t$} \label{line:taurstep}
    \STATE{$\one./\taubf_p^{\tp1} \gets \theta^t\one./\taubfbar_p^{\tp1}
        + (1-\theta^t)\one./\taubf_p^{t}$} \label{line:taupstep}

\UNTIL{Terminated}
\end{algorithmic}
\end{algorithm}

Another point to note in reading Algorithm~\ref{algo:ALG}
is that the expectation and variance operators in \eqref{eq:JlinBFE}, \eqref{eq:ADMMs}, and \eqref{eq:ADMMq} have been replaced by
componentwise estimation functions $g_x$ and $g_z$ and their scaled derivatives.
In particular, recall from \eqref{eq:bxz} that $b_x^{\tp1}$ is fully parameterized by $(\rbf^t,\taubf_r^t)$ and that $b_z^{\tp1}$ is fully parameterized by $(\pbf^t,\taubf_p^t)$.
Thus, we can write the means of these distributions as
\begin{subequations}
\label{eq:GxzSP}
\beqa
    \xbf^{\tp1}
    &=& \Exp(\xbf|b_x^{\tp1})
    \defn g_x(\rbf^t,\taubf_r^t),
    \\
    \zbf^{\tp1}
    &=& \Exp(\zbf|b_z^{\tp1})
    \defn g_z(\pbf^t,\taubf_p^t) ,
\eeqa
\end{subequations}
as reflected in line~\ref{line:xzest} of Algorithm~\ref{algo:ALG}.
For separable $\fx$ and $\fz$, we note that the computations in \eqref{eq:GxzSP} can be performed in a componentwise, scalar manner, e.g.,
\beqa
    x_j^{\tp1}
    &=& [g_{x}(\rbf^t,\taubf_r^t)]_j \defn g_{x_j}(r_j^t,\tau_{r_j}^t) \nonumber \\
    &=& \frac
    {\int_{\R} x \exp\big( -\fxj(x)-\frac{1}{2\tau_{r_j}^t}(x-r_j^t)^2\big) \d x}
    {\int_{\R} \exp\big( -\fxj(x)-\frac{1}{2\tau_{r_j}^t}(x-r_j^t)^2\big) \d x}
    \label{eq:gxSP}, \\
    z_i^{\tp1}
    &=& [g_{z}(\pbf^t,\taubf_p^t)]_j \defn g_{z_i}(p_i^t,\tau_{p_i}^t) \nonumber \\
    &=& \frac
    {\int_{\R} z \exp\big( -\fzi(z)-\frac{1}{2\tau_{p_i}^t}(z-p_i^t)^2\big) \d z}
    {\int_{\R} \exp\big( -\fzi(z)-\frac{1}{2\tau_{p_i}^t}(z-p_i^t)^2\big) \d z}
    \label{eq:gzSP}, 
\eeqa
Furthermore, the variances of $b_{x_j}^{\tp1}$ and $b_{z_i}^{\tp1}$ can be computed in a componentwise manner using the derivatives of $g_{x_j}$ and $g_{z_i}$ with respect to their first argument \cite{Rangan:11-ISIT}, i.e.,
\begin{subequations}
\beqa
    \taubf_x^{\tp1}
    &\defn& \var(\xbf|b_x^{\tp1})
    = \taubf_r^t . g_x'(\rbf^t,\taubf_r^t) ,
    \\
    \taubf_z^{\tp1}
    &\defn& \var(\zbf|b_z^{\tp1})
    = \taubf_p^t . g_z'(\pbf^t,\taubf_p^t) ,
\eeqa
\end{subequations}
as reflected in line~\ref{line:tauxzt} of Algorithm~\ref{algo:ALG}.
That is,
\beqa
    \tau_{x_j}^{\tp1}
    &=& [\taubf_r^t . g_x'(\rbf^t,\taubf_r^t)]_j
    = \tau_{r_j}^t \frac{\partial g_{x_j}(r_j^t,\tau_{r_j}^t)}{\partial r_j}
    \label{eq:tauxSP} .
\eeqa
We use these general scalar estimation functions $g_x$
and $g_z$ since it will allow us later to consider
a similar algorithm for the MAP estimation problem \eqref{eq:xhatMAP}.

Interestingly, the \ALG algorithm has close similarities to the sum-product version of the
original GAMP algorithm from \cite{Rangan:11-ISIT}, as we will discuss in Section~\ref{sec:gamp}.
For example, the sum-product version of the GAMP algorithm uses the same estimation functions $g_x$ and $g_z$ from \eqref{eq:GxzSP},
which we will refer to as the \emph{\SP estimation functions}.

\color{black}
\subsection{Computational Cost}

While we will demonstrate below that \ALG offers improved convergence stability
relative to the GAMP algorithm of \cite{Rangan:11-ISIT}, 
it is important to point out that the computational cost of \ALG may be somewhat larger:
One of the main attractive features 
of GAMP and  other first order methods, is that each iteration
requires only matrix-vector multiplies by by $\Abf$ and $\Abf\tran$.
Each such multiplication will have complexity $O(mn)$ in the most general case,
and  may be smaller for structured transforms such as filters, FFTs, or sparse matrices.

In contrast, \ALG requires a least-squares (LS) minimization \eqref{eq:vmin} in each iteration.  
Exact evaluation of the minimization will have a cost of $O(n^2m)$ -- a cost not incurred
in GAMP or most other first-order methods.  As is done 
ADMM \cite{BoydPCPE:09} -- and in the simulations below --  the minimization
can be performed approximately via conjugate gradient (CG) \cite{nocedal2006numerical}.
Conjugate gradient also requires repeated matrix-vector multiplies by 
$\Abf$ and $\Abf\tran$, but will require $K$ such matrix-vector multiplies
where $K$ is the number of CG iterations.  In the simulations below, we will use $K=3$,
thus increasing the per iteration cost of \ALG by a factor of approximately 
3 relative to standard GAMP.  

The other computations in each iteration of \ALG are typically smaller than
the LS minimization and are similar to those performed in GAMP.
For example, similar to GAMP, each iteration requires evaluation of the estimation functions
$g_x(\cdot)$ and $g_z(\cdot)$.  These can be performed
as $n$ and $m$ componentwise scalar functions given in \eqref{eq:gxSP} and \eqref{eq:gzSP}. 
For certain penalty functions, such as Bernoulli-Gaussians, these will have closed-form
expressions; otherwise, they will need to be evaluated via numerical integration.
In either case, the componentwise cost does not grow with the dimension, so the per iteration
cost of evaluating the estimation functions is $O(m+n)$ and are typically
not dominant for large $m$ and $n$.
\color{black}


\section{\ALG for MAP Estimation} \label{sec:map}

\subsection{ADMM Inner Loop}
For the posterior density $p(\xbf|\ybf)$ in \eqref{eq:pxy},
the MAP estimates of the vector $\xbf$ and its transform
$\zbf =\Abf\xbf$ are given by the constrained optimization
\beq \label{eq:xzmap}
    (\xbfhat,\zbfhat) \defn \argmin_{\xbf,\zbf} J(\xbf,\zbf) \mbox{ s.t. }
        \zbf = \Abf\xbf,
\eeq
where $J(\xbf,\zbf)$ is the objective function
\beq \label{eq:JMAP1}
    J(\xbf,\zbf) \defn \fx(\xbf) + \fz(\zbf).
\eeq
We will show that, with appropriate selection
of the estimation functions, $g_x$ and $g_z$,
the inner loop of
Algorithm~\ref{algo:ALG} can be used as an ADMM method for solving
\eqref{eq:xzmap}.

As before, we replace the constraint $\zbf=\Abf\xbf$
in the optimization \eqref{eq:xzmap} with
two constraints: $\xbf=\vbf$ and $\zbf = \Abf\vbf$.
We then define the augmented Lagrangian
\beqa
    \lefteqn{ L(\xbf,\zbf,\sbf,\qbf,\vbf;\taubf_p,\taubf_r) }
    \nonumber\\
    &\defn& \fx(\xbf) + \fz(\zbf)
    + \qbf\tran(\xbf-\vbf)
    + \sbf\tran(\zbf-\Abf \vbf)
    \nonumber \\ && \mbox{}
    + \tfrac{1}{2}\|\xbf-\vbf\|^2_{\taubf_r}
    + \tfrac{1}{2}\|\zbf-\Abf\vbf\|^2_{\taubf_p}.
    \label{eq:LaugMS}
\eeqa
The ADMM recursions \eqref{eq:admmGen} for this augmented
Lagrangian are then given by
\begin{subequations} \label{eq:ADMMmap}
\beqa
    (\xbf^{\tp1},\zbf^{\tp1})
    &=& \argmin_{\xbf,\zbf}
        L(\xbf,\zbf,\sbf^t,\qbf^t,\vbf^t;\taubf_p,\taubf_r),
	\label{eq:ADMMmapxz}\\
    \sbf^{\tp1}
    &=& \sbf^{t} + \Diag(\one./\taubf_p)\big(\zbf^{\tp1}-\Abf\vbf^{t}\big),
    	\label{eq:ADMMmaps}\\
    \qbf^{\tp1}
    &=& \qbf^{t} + \Diag(\one./\taubf_r)\big(\xbf^{\tp1}-\vbf^{t}\big),
    	\label{eq:ADMMmapq}\\
    \vbf^{\tp1}
    &=& \argmin_{\vbf} L(\xbf^{\tp1},\zbf^{\tp1},\sbf^{\tp1},\qbf^{\tp1},\vbf;\taubf_p,\taubf_r) . \quad\qquad
	\label{eq:ADMMmapv}
\eeqa
\end{subequations}
To perform the minimization in \eqref{eq:ADMMmapxz},
first consider the minimization over $\xbf$.
Eliminating terms that do not depend on $\xbf$, we obtain
\beqa
    \xbf^{\tp1} &=& \argmin_{\xbf}
        \fx(\xbf) + \qbf\tran\xbf + \tfrac{1}{2}\|\xbf-\vbf\|^2_{\taubf_r} \nonumber\\
    &=& \argmin_{\xbf} \fx(\xbf) +
        \tfrac{1}{2}\|\xbf+\taubf_r.\qbf-\vbf\|^2_{\taubf_r}.
            \label{eq:ADMMmapx}
\eeqa
Similarly, the minimization over $\zbf$ reduces to
\beqa
    \zbf^{\tp1} &=& \argmin_{\zbf}
        \fz(\zbf) + \sbf\tran\zbf +
        \tfrac{1}{2}\|\zbf-\Abf\vbf\|^2_{\taubf_p} \nonumber\\
    &=& \argmin_{\zbf} \fz(\zbf) +
        \tfrac{1}{2}\|\zbf+\taubf_p.\sbf-\Abf\vbf\|^2_{\taubf_p}.
            \label{eq:ADMMmapz}
\eeqa
Hence, if we define the estimation functions
\begin{subequations} \label{eq:GxzMS}
\beqa
    g_x(\rbf,\taubf_r) &\defn& \argmin_{\xbf} \left[
    \fx(\xbf) +
        \tfrac{1}{2}\|\xbf-\rbf\|^2_{\taubf_r}\right], \\
    g_z(\pbf,\taubf_p) &\defn& \argmin_{\zbf} \left[
    \fx(\xbf) +
        \tfrac{1}{2}\|\zbf-\pbf\|^2_{\taubf_p}\right],
\eeqa
\end{subequations}
then we can rewrite \eqref{eq:ADMMmapx} and \eqref{eq:ADMMmapz}
as
\beq \label{eq:xzgMS}
    \xbf^{\tp1} = g_x(\rbf^t,\taubf_r), \quad
    \zbf^{\tp1} = g_z(\pbf^t,\taubf_p),
\eeq
for $\rbf^t$ and $\pbf^t$ defined in \eqref{eq:rpt}.
Also, the minimization over $\vbf$ in \eqref{eq:ADMMmapv}
can again be cast as the least-squares problem \eqref{eq:vmin}.

We see that equations \eqref{eq:rpt}, \eqref{eq:vmin},
\eqref{eq:ADMMmaps},
\eqref{eq:ADMMmapq} and \eqref{eq:xzgMS} are precisely the updates in the
ADMM inner-loop of Algorithm~\ref{algo:ALG}.  Therefore, for fixed
penalty terms $\taubf_r$ and $\taubf_p$, the inner loop of the
\ALG algorithm with the estimation functions \eqref{eq:GxzMS}
is precisely an ADMM algorithm for the MAP estimation
problem~\eqref{eq:xzmap}.

The functions in \eqref{eq:GxzMS} are the standard
``proximal" operators used in many implementations of ADMM and related
optimization algorithms \cite{BoydPCPE:09}.
These functions also appear in the max-sum version of GAMP from \cite{Rangan:11-ISIT}, which is used for MAP estimation.
Thus, we will refer to \eqref{eq:GxzMS} as the \emph{\MS estimation functions}.

\subsection{Hardening Limit of the LSL-BFE} \label{sec:hardMap}

The above discussion shows that, with the \MS estimation
functions \eqref{eq:GxzMS}, the \ALG outputs
$(\xbf^t,\zbf^t)$ can be interpreted as estimates of the
MAP solution from~\eqref{eq:xzmap}.
How then do we interpret the related
terms $(\taubf_x^t,\taubf_z^t)$?  In the inference (i.e., MMSE) problem
from Section~\ref{sec:mmse}, the components of $\taubf_x^t$ and $\taubf_z^t$
are estimates of the variances of the marginal posteriors.
Below, we use a hardening argument to show that, in the MAP problem,
$(\taubf_x^t,\taubf_z^t)$ can be interpreted as estimates of the local
curvature of the MAP objective \eqref{eq:JMAP1}.

To be precise, let us define the
\emph{marginal minimization functions}
\begin{subequations} \label{eq:margMin}
\beqa
    \phi_{x_j}(x_j)
    &\defn&  \min_{\xbf \setminus x_j} J(\xbf,\Abf\xbf),\\
    \phi_{z_i}(z_i)
    &\defn& \min_{\xbf: z_i=[\Abf\xbf]_i} J(\xbf,\Abf\xbf),
\eeqa
\end{subequations}
where the minimizations are over the vector $\xbf$, holding either $x_j$
or $z_i \defn [\Abf\xbf]_i$ fixed.
Note that, if one can compute these marginal minimization functions,
then one can compute the components of the MAP estimates from \eqref{eq:xzmap}
via
\beq \label{eq:xzmapPhi}
    \xhat_j = \argmin_{x_j} \phi_{x_j}(x_j), \quad
    \zhat_i = \argmin_{z_i} \phi_{z_i}(z_i).
\eeq
However, the marginal minimization functions provide not only
componentwise objectives for the MAP optimization \eqref{eq:xzmap},
but also the sensitivity of those objectives.


We will see that \ALG provides estimates of the marginal minimization
functions, in addition to estimates of the MAP solution in \eqref{eq:xzmap}.
Perhaps the easiest way to see this is through a standard ``hardening"
analysis, which is also used to understand how max-sum
loopy belief propagation can be viewed as a limit of sum-product loopy BP;
see, for example, \cite{weiss2007map,RanganFG:12-IT}.
Specifically, combining \eqref{eq:pxy} with
Laplace's Principle from large deviations \cite{DemboZ:98},
and assuming suitable continuity conditions, the marginal
minimization functions \eqref{eq:margMin} are given by
(up to a constant factor)
\beqan
    && \phi_{x_j}(x_j) = - \lim_{T \arr 0} T \ln \pdf_{x_j}(x_j;T), \\
    && \phi_{z_i}(z_i) = - \lim_{T \arr 0} T \ln \pdf_{z_i}(z_i;T),
\eeqan
where $\pdf_{x_j}(x_j;T)$ and $\pdf_{z_i}(z_i;T)$ are the marginal densities
for the \emph{scaled} joint density
\[
    \pdf(\xbf;T) \defn \frac{1}{Z}\exp\left[ -\frac{1}{T}\Big(\fx(\xbf) + \fz(\Abf\xbf)\Big) \right] .
\]
Note that, for any $T > 0$, we can estimate the marginal posteriors $\pdf_{x_j}(x_j;T)$ and $\pdf_{z_i}(z_i;T)$
using the LSL-BFE optimization from Section~\ref{sec:mmse}.
That is, we can use the estimate
\begin{subequations} \label{eq:phiLimT}
\beqa
    \phi_{x_j}(x_j) & \approx & \phihat_{x_j}(x_j) \defn
          - \lim_{T \arr 0} T \ln \bhat_{x_j}(x_j;T), \\
    \phi_{z_i}(z_i) & \approx & \phihat_{z_i}(z_i) \defn
            - \lim_{T \arr 0} T \ln \bhat_{z_i}(z_i;T),
\eeqa
\end{subequations}
where $\bhat_{x_j}(x_j;T)$ and $\bhat_{z_i}(z_i;T)$
are the belief estimates
computed via the LSL-BFE optimization under the scaled penalties
\beq \label{eq:fxzScale}
    \fx(\xbf;T) \defn \fx(\xbf)/T, \quad \fz(\zbf;T) \defn \fz(\zbf)/T.
\eeq
In statistical physics, the parameter $T$ has the interpretation of temperature,
and the limit $T \arr 0$ corresponds to a ``cooling" of the system.
In inference problems, the cooling has the effect of concentrating the distributions
about their maxima.

A large-deviations analysis in Appendix~\ref{sec:largeDev}
shows that, if we use \ALG with the \SP estimation functions
\eqref{eq:GxzSP} with the scaled functions \eqref{eq:fxzScale},
then at iteration $t$ the limits in \eqref{eq:phiLimT} are given by
\begin{subequations} \label{eq:phihat}
\beqa
    \phihat_{x_j}^t(x_j) &=& -\lim_{T \arr 0} T\ln b^t_{x_j}(x_j;T),
   \nonumber\\
   &=&
        \fxj(x_j) + \tfrac{1}{2\tau_{r_j}^t}(x_j-r_j^t)^2 \\
    \phihat_{z_i}^t(z_i) &=& -\lim_{T \arr 0} T\ln b^t_{z_i}(z_i;T)
    \nonumber \\
        &=& \fzi(z_i) + \tfrac{1}{2\tau_{p_i}^t}(z_i-p_i^t)^2,
\eeqa
\end{subequations}
where the parameters $r_j^t$, $p_i^t$, $\tau_{r_j}^t$, and $\tau_{p_i}^t$
are the outputs of \ALG under the \MS estimation functions
\eqref{eq:GxzMS}.
In this sense, \ALG under the \MS estimation function can be seen as a limiting case of \ALG under the \SP estimation functions.
Hence, according to \eqref{eq:phiLimT},
\MS \ALG can be used to compute estimates \eqref{eq:phihat}
of the marginal minimization functions \eqref{eq:margMin}.
Furthermore, according to \eqref{eq:GxzMS} and \eqref{eq:xzgMS},
$\xbf^{\tp1}$ and $\zbf^{\tp1}$ are the minima of these functions
\[
    \xhat^{\tp1}_j = \argmin_{x_j} \phihat_{x_j}^t(x_j), \quad
    \zhat^{\tp1}_i = \argmin_{z_i} \phihat_{z_i}^t(z_i),
\]
as one would expect from \eqref{eq:xzmapPhi}.

Finally, it can be shown (see \eqref{eq:gxderivMS}) that, for the \MS
estimation functions \eqref{eq:GxzMS}, the outputs of
line~\ref{line:tauxzt} in Algorithm~\ref{algo:ALG} take the form
\begin{subequations}
\beqa
    \tau_{x_j} &=& \tau_{r_j}g_{x_j}'(r_j,\tau_{p_i})
        = \frac{\tau_{r_j}}{1 + \tau_{r_j} f_{x_j}''(\xhat_j)},  \\
    \tau_{z_i} &=& \tau_{p_i}g_{z_i}'(p_i,\tau_{p_i})
        = \frac{\tau_{p_i}}{1 + \tau_{p_i} f_{z_i}''(\zhat_i)}.
\eeqa
\end{subequations}
Meanwhile, from \eqref{eq:phihat}, we see that
\beq \label{eq:tauxzMAP}
    \frac{1}{\tau_{x_j}^{\tp1}} =
        \frac{\partial^2\phihat_{x_j}^t(\xhat_j^{\tp1})}{\partial x_j^2}, \quad
    \frac{1}{\tau_{z_i}^{\tp1}} =
        \frac{\partial^2\phihat_{z_i}^t(\zhat_i^{\tp1})}{\partial z_i^2}.
\eeq
Therefore, when \ALG is used for MAP estimation,
the components of $\taubf_x^t$ and $\taubf_z^t$ can be
interpreted as the inverse curvatures of the constrained
function $J(\xbf,\zbf=\Abf\xbf)$ in the vicinity of the current estimate $(\xbfhat^t,\zbfhat^t)$.

Appendix~\ref{sec:largeDev} also show that, in the limit
as $T \arr 0$, the LSL-BFE optimization \eqref{eq:FbetheOpt} decomposes
approximately into two decoupled optimizations:
The first computes the MAP estimates $(\xbfhat,\zbfhat)$ from \eqref{eq:xzmap},
and the second computes
\beq \label{eq:JMAP2Opt}
    (\taubfhat_x,\taubfhat_z) \defn \argmin_{\taubf_x,\taubf_z}
        J^2(\taubf_x,\taubf_z,\xbfhat,\zbfhat),
\eeq
where
\beqa \label{eq:JMAP2}
    J^2(\taubf_x,\taubf_z,\xbfhat,\zbfhat)
    &\defn& \sum_{j=1}^n  \left[ \tau_{x_j}\fxj''(\xhat_j)
        - \ln(\tau_{x_j}) \right]
        \\ &&
        + \sum_{i=1}^m \left[ \tau_{z_i}\left( \fzi''(\zhat_i) + \frac{1}{\tau_{p_i}}\right)
            + \ln\left(\frac{\tau_{p_i}}{\tau_{z_i}}\right) \right],
            \nonumber
\eeqa
and, as before, $\taubf_p \defn \Sbf\taubf_x$.
Since the optimization \eqref{eq:JMAP2Opt} provides the inverse-curvature
estimates in \eqref{eq:tauxzMAP}, we will refer to it as
\emph{curvature optimization}.

\section{Convergence Analysis for Strictly Convex Penalties}

\subsection{Fixed Points of \ALG}
We first characterize the fixed points of \ALG,
assuming that the algorithm converges.

\medskip
\begin{theorem} \label{thm:fix}  At any fixed point
of \ALG with the \SP estimation functions \eqref{eq:GxzSP},
the belief pair $(b_x,b_z)$ in \eqref{eq:bxz}
is a critical point of the constrained LSL-BFE optimization
\eqref{eq:FbetheOpt}.
\end{theorem}
\begin{IEEEproof}  See Appendix~\ref{sec:fixPf}.
\end{IEEEproof}

\begin{theorem} \label{thm:fixMS}  At any fixed point
of \ALG with the \MS estimation functions \eqref{eq:GxzMS},
the output $(\xbf,\zbf)$ is a critical point of the constrained
MAP optimization \eqref{eq:xzmap}
and $(\taubf_x,\taubf_z)$ is a critical point of the
optimization \eqref{eq:JMAP2Opt}.
\end{theorem}
\begin{IEEEproof}  See Appendix~\ref{sec:fixPf}.
\end{IEEEproof}

\medskip
Theorems~\ref{thm:fix} and \ref{thm:fixMS}
show that, if \ALG converges, then
its limit points will be local minima of
either the inference (i.e., MMSE) or MAP problems.

\subsection{Convergence of the ADMM Inner Loop} \label{sec:meanConv}


For the remainder of this section, we will show the convergence of
ADMM-GAMP in the special case of convex and smooth penalties
$f_x$ and $f_z$. We begin by analyzing the convergence of the ADMM inner-loop
under fixed linearization terms $\taubf_r$ and $\taubf_p$.
It is well-known that, when one applies ADMM
to a general optimization problem of the form \eqref{eq:optGen}
with convex $f$ and full-rank $\Bbf$, the method will converge
\cite{BoydPCPE:09}.
However, in our case, the objective function is
the linearized LSL-BFE in \eqref{eq:JlinBFE}, which is not necessarily
convex, even if the penalty functions $\fx$ and $\fz$
are.  The problem is that
the variances $\var(\xbf|b_x)$ and $\var(\zbf|b_z)$
are not convex functions of the densities $b_x$ and $b_z$
(in fact, they are concave).  We thus need a separate proof.

We will prove convergence under the following assumption.

\medskip

\begin{assumption}  \label{as:cont}
For fixed $\taubf_r$ and $\taubf_p$,
the estimation functions $g_x(\rbf,\taubf_r)$ and $g_z(\pbf,\taubf_p)$ are separable in $\rbf$ and $\pbf$ in that
\beqan
    g_x(\rbf,\taubf_r) &=& \big(g_{x_1}(r_1,\taubf_r), \cdots, g_{x_n}(r_n,\taubf_r) \big), \\
    g_z(\pbf,\taubf_p) &=& \big(g_{z_1}(p_1,\taubf_p), \cdots, g_{z_m}(p_m,\taubf_p) \big)
\eeqan
for scalar function $g_{x_j}$ and $g_{z_i}$.
In addition, these scalar functions
have, with respect to their first arguments, continuous first derivatives $g_{x_j}'$ and $g_{z_i}'$ satisfying
\beq \label{eq:gxzdbnd}
    \epsilon \leq g_{x_j}'(r_j,\taubf_r) \leq 1-\epsilon, \quad
    \epsilon \leq g_{z_i}'(p_i,\taubf_p) \leq 1-\epsilon.
\eeq
for some constant $\epsilon\in (0,0.5]$.
\end{assumption}

The assumption requires that the estimation functions are strictly increasing contractions.
Importantly, the following lemma shows that this assumption holds when the penalty functions are smooth and convex.

\begin{lemma} \label{lem:gspConv}
Suppose that $\fx$ and $\fz$ are strictly convex, separable functions, in that they are of the
form \eqref{eq:fxzsep}, where the components have continuous second derivatives such that
\beq \label{eq:fxzderivBnd}
    A \leq \fxj''(x_j) \leq B~\forall x_j, \quad A \leq \fzi''(z_i) \leq B~\forall z_i,
\eeq
for some $0 < A \leq B < \infty$.  Then, both the \SP
estimation functions in \eqref{eq:GxzSP}
and the \MS estimation functions in \eqref{eq:GxzMS}
satisfy Assumption~\ref{as:cont} for any $\taubf_r,\taubf_p > 0$.
\end{lemma}
\begin{IEEEproof} See Appendix \ref{sec:gspConvPf}.
\end{IEEEproof}

We now have the following convergence result.

\medskip
\begin{theorem} \label{thm:meanConv}
Consider Algorithm~\ref{algo:ALG} with only ADMM updates
(i.e., $\theta^t=0$ for all $t$), so that the linearization terms remain
constant, i.e.,
\[
    \taubf_p^t = \taubf_p \text{~and~} \taubf_r^t = \taubf_r ~\forall t,
\]
for some vectors $\taubf_p$ and $\taubf_r$.
Then, if the estimation functions satisfy Assumption~\ref{as:cont},
the algorithm converges to a unique fixed point at a linear rate of convergence
of $1-\epsilon$.
\end{theorem}
\begin{IEEEproof} See Appendix \ref{sec:meanConvPf}.
\end{IEEEproof}

\subsection{Outer Loop Convergence:  MMSE Case} \label{sec:varConv}

Theorem~\ref{thm:meanConv} shows that, with the \SP estimation
functions \eqref{eq:GxzSP} and strictly convex penalties,
the ADMM inner loop of Algorithm~\ref{algo:ALG} converges.
We next consider the convergence of the outer loop,
Algorithm~\ref{algo:outer}, assuming that the inner minimization
(i.e., line~\ref{line:lslMin} of Algorithm~\ref{algo:outer})
is computed exactly.
\medskip

\begin{theorem} \label{thm:varConv}
Suppose that the functions $\fx$ and $\fz$
satisfy the assumptions in Lemma~\ref{lem:gspConv}
and the matrix $\Sbf$ has positive components
(i.e., $S_{ij}=|A_{ij}|^2 > 0~\forall ij$).
Then, there exists a $\thetabar$ such that, if $\theta^k <\thetabar$,
the sequence of belief estimates $b^k$ generated by
Algorithm~\ref{algo:outer} yields a monotonically non-increasing LSL-BFE,
i.e.,
\[
    J(b^{\kp1}_x,b^{\kp1}_z) \leq J(b^k_x,b^k_z).
\]
\end{theorem}
\begin{IEEEproof} See Appendix \ref{sec:varConvPf}.
\end{IEEEproof}

\medskip
Together, Theorems \ref{thm:meanConv} and \ref{thm:varConv} demonstrate
that \ALG will converge under an infinitely slow
damping schedule.  Specifically, we select iterations $t_1 < t_2 < \cdots$
that are infinitely far apart.
Then, for all $t$ between each $t_k$ and $t_{\kp1}$,
we set $\theta^t=0$ so that the ADMM inner-loop is run to completion,
and at each $t=t_k$, we select $\theta^t$ to be a small positive value.



It is of course impossible to use an infinite number of
inner-loop iterations in practice.
Fortunately, our numerical experiments in Section~\ref{sec:Simulations}
suggest that a fixed number of inner-loop iterations is sufficient.

\subsection{Outer Loop Convergence:  MAP Case} \label{sec:varConvMS}

We can prove a stronger convergence result for \ALG under the \MS
estimation functions \eqref{eq:GxzMS}, if we make two additional assumptions.
Recall from Theorem~\ref{thm:meanConv} that, if we
set $\theta^t=0~\forall t$, then
the linearization parameters $\taubf_r^t$ and $\taubf_p^t$ will remain
constant with $t$ and the algorithm will converge to some fixed point.
Our first assumption is that we begin the algorithm at one such fixed point.
That is, we suppose that the time $t=0$ versions of
\beq \label{eq:initVarMS}
    \xbf^t,~\zbf^t,~\rbf^t,~\pbf^t,~\qbf^t,~\sbf^t,~\vbf^t
\eeq
are fixed points of lines \ref{line:rt} through \ref{line:vt}
in Algorithm~\ref{algo:ALG}.
Our second assumption is that we replace the $\taubf_s^{\tp1}$
update in line~\ref{line:taust} with
\beq \label{eq:taustMS}
    \taubf_s^{\tp1} \gets
        (\one - \taubf_z^{\tp1}./\taubf_p^t)./\taubf_p^t.
\eeq
That is, we use $\taubf_p^t$ instead of $\taubfbar_p^{\tp1}$.
Under these two additional assumptions, we can prove the following.

\medskip
\begin{theorem} \label{thm:varConvMS}
Consider \ALG, Algorithm~\ref{algo:ALG}, run under the \MS
estimation functions \eqref{eq:GxzMS}, with penalty functions $\fx$
and $\fz$ satisfying the assumptions of Lemma~\ref{lem:gspConv}.
Suppose that the initialization \eqref{eq:initVarMS} is a
fixed point of lines \ref{line:rt} through \ref{line:vt},
and that line~\ref{line:taust} is replaced by \eqref{eq:taustMS}.
Then, if $\theta^t=1$ for all $t$,
\begin{enumerate}[(a)]
\item Even though $\taubf_r^t$ and $\taubf_p^t$ may change with $t$,
the variables in \eqref{eq:initVarMS}
will remain constant.  That is, for all $t$,
\beqa \label{eq:initVarConst}
    && \xbf^t = \xbf^0, ~ \zbf^t = \zbf^0, ~ \rbf^t = \rbf^0,
        ~ \pbf^t = \pbf^0, \nonumber \\
    && \qbf^t = \qbf^0, ~\sbf^t = \sbf^0, ~\vbf^t = \vbf^0.
\eeqa
Moreover, the variables $(\xbf^0,\zbf^0)$ are the global minima
of the MAP estimation problem \eqref{eq:xzmap}.
\item The linearization parameters $\taubf_x^t$ and $\taubf_z^t$
converge to unique global minima of the curvature optimization
\eqref{eq:JMAP2Opt}.
\end{enumerate}
\end{theorem}
\begin{IEEEproof}  See Appendix~\ref{sec:varConvMSPf}.
\end{IEEEproof}

\medskip
The result shows that, in principle, we can solve the MAP estimation
problem by first running the ADMM inner loop to convergence with
arbitrary positive linearization terms $\taubf_r$ and $\taubf_p$.
Then, we could turn on the
outer loop updates, thus driving $\taubf_x^t$ and $\taubf_z^t$ to
the minima of the curvature optimization problem \eqref{eq:tauxzMAP}.
Of course, in practice, one cannot do this perfectly,
since the ADMM inner loop must be terminated at some finite number
of iterations.
Also, it is possible that, by letting the variance terms adapt
(at least slowly) before the inner loop fully converges, the
convergence speed of the inner loop can be improved.
In fact, this is our empirical experience, although we have no proof.

It is important to point out that the MAP convergence proof
requires a slightly modified variance update given in
\eqref{eq:taustMS}.  This update may actually be preferable for the
MMSE case as well, however, further analysis would be required.
Indeed, while we have demonstrated one variance update with provable
convergence, finding the best variance update method is a still an open question.

\section{Relationship of \ALG to GAMP}	\label{sec:gamp}

There are two key differences between the proposed \ALG algorithm and the original sum-product GAMP algorithm from \cite{Rangan:11-ISIT}, reproduced for convenience in Algorithm~\ref{algo:GAMP} (with the variance updates indented for visual clarity).
\begin{enumerate}
\item
The \ALG algorithm uses two additional variables: a dual variable $\qbf^t$, and an auxiliary variable $\vbf^t$ that is updated via the least-squares optimization \eqref{eq:vmin}, that are not present in the original GAMP algorithm.
\item
\ALG uses an alternating schedule of mean and (possibly damped) variance updates, whereas GAMP uses interleaved mean and variance updates.
\end{enumerate}
Below, we describe these differences in more detail.

\begin{algorithm}
\caption{Original GAMP}
\begin{algorithmic}[1]  \label{algo:GAMP}
\REQUIRE{ Matrix $\Abf$ and estimation functions $g_x$ and $g_z$. }

\STATE{ $\Sbf \gets \Abf.\Abf$ (componentwise square) }
\STATE{ Initialize $\xbf^0$, $\taubf_x^0$  }
\STATE{ $\sbf^{0} \gets 0$ }
\STATE{ $t \gets 0$  }
\REPEAT

    \STATE{ ~~~$\taubf_p^{t} \gets \Sbf\taubf_x^{t}$ }
        \label{line:taupSP}
    \STATE{ $\pbf^t \gets \Abf\xbf^t -
        \taubf_p^t.\sbf^{\tm1}$ } \label{line:phatSP}
    \STATE{ ~~~$\taubf_z^t \gets
	\taubf_p^t. g_z'(\pbf^t,\taubf_p^t)$} \label{line:tauzSP}
    \STATE{ $\zbf^t \gets g_z(\pbf^t,\taubf_p^t)$} \label{line:zhatSP}
    \STATE{ ~~~$\taubf_s^t \gets (\one- \taubf_z^t./\taubf_p^t)./\taubf_p^t$ }
           \label{line:tausSP}
    \STATE{ $\sbf^t \gets (\zbf^t-\pbf^t)./\taubf_p^t$ }
            \label{line:shatSP}
    \STATE{ ~~~$\taubf_r^t \gets \one./(\Sbf\tran\taubf_s^t)$ }
            \label{line:taurSP}
    \STATE{ $\rbf^t \gets \xbf^t + \Diag(\taubf_r^t)\Abf\tran\sbf^t$ }
    \STATE{ ~~~$\taubf_x^{\tp1} \gets \taubf_r^t .g_x'(\rbf^t,\taubf_r^t)$}
            \label{line:tauxSP}
    \STATE{ $\xbf^{\tp1} \gets g_x(\rbf^t,\taubf_r^t)$}
            \label{line:xhatSP}

\UNTIL{Terminated}

\end{algorithmic}
\end{algorithm}

\subsection{Sum-product GAMP via Stale, Linearized ADMM}

One way to understand the differences between \ALG and the original GAMP is as follows:
\ALG results from minimizing the linearized LSL-BFE via ADMM under the splitting rule ``$\Exp(\zbf|b_z)=\Abf\vbf$ and $\Exp(\xbf|b_x)=\vbf$" (as described in Section~\ref{sec:admmBFE1}), whereas the original GAMP uses \emph{stale, linearized} ADMM under the conventional\footnote{See, e.g., \cite[Sec.~3.1]{BoydPCPE:09}.} splitting rule ``$\Exp(\zbf|b_z)=\Abf\Exp(\xbf|b_x)$."
Both use the same iterative LSL-BFE linearization strategy described in Section~\ref{sec:milBFE}.

We can derive the mean updates in the original GAMP using the augmented Lagrangian
\beqa
    L(b_x,b_z,\sbf;\taubf_p)
    &\defn& J(b_x,b_z,\taubf_r,\taubf_p)
    + \sbf\tran\big(\Exp(\zbf|b_z)-\Abf \Exp(\xbf|b_x)\big)
    \nonumber \\ && \mbox{}
    + \tfrac{1}{2}\|\Exp(\zbf|b_z)-\Abf\Exp(\xbf|b_x)\|^2_{\taubf_p}
    \label{eq:LaugGAMP} ,
\eeqa
for the $J$ defined in 
\eqref{eq:JlinBFE} and \emph{stale, linearized} ADMM:
\begin{subequations} \label{eq:LADMM}
\beqa
    b_x^{\tp1}
    &=& \argmin_{b_x} L(b_x,b_z^t,\sbf^{\tm1};\taubf_p)
    	+ \tfrac{1}{2}\big(\Exp(\xbf|b_x)-\Exp(\xbf|b_x^t)\big)\tran
        \nonumber \\ && \mbox{} \times
	\big(\Dbf_{\taubf_r}-\Abf\tran\Dbf_{\taubf_p}\Abf\big)
	\big(\Exp(\xbf|b_x)-\Exp(\xbf|b_x^t)\big),
	\label{eq:LADMMx}\\
    b_z^{\tp1}
    &=& \argmin_{b_z} L(b_x^{\tp1},b_z,\sbf^t;\taubf_p),
	\label{eq:LADMMz}\\
    \sbf^{\tp1}
    &=& \sbf^{t} + \Dbf_{\taubf_p}\big(\Exp(\zbf|b_z^{\tp1})-\Abf\Exp(\xbf|b_x^{\tp1})\big)
    	\label{eq:LADMMs} ,
\eeqa
\end{subequations}
where $\Dbf_{\taubf}\defn\Diag(\one./\taubf)$.
Note the addition of a ``linearization" term in \eqref{eq:LADMMx} to decouple the minimization.
The resulting approach goes by several names: linearized ADMM \cite[Sec.~4.4.2]{ParikhB:13}, split inexact Uzawa \cite{Esser:JIS:10}, and primal-dual hybrid gradient (PDHG) \cite{Esser:JIS:10}.
Note also the use of the ``stale" dual estimate $\sbf^{\tm1}$ in \eqref{eq:LADMMx}, as opposed to the most recent dual estimate $\sbf^t$.
In the context of PDHG, this stale update is known as Arrow-Hurwicz \cite{Esser:JIS:10}.
In Appendix~\ref{sec:original}, we show that the recursion \eqref{eq:LADMM} yields the mean updates in the original sum-product GAMP algorithm (i.e., the non-indented lines in Algorithm~\ref{algo:GAMP}).

Regarding the variance updates of the original sum-product GAMP algorithm (i.e., the indented lines in Algorithm~\ref{algo:GAMP}), a visual inspection shows that they match the non-damped \ALG ``gradient" updates (i.e., lines~\ref{line:tauxzt}-\ref{line:taurbart} of Algorithm~\ref{algo:ALG} under $\theta^t=1$), except for one small difference: in the original sum-product GAMP, the update of $\taubf_s$ uses the same version of $\taubf_p$ used by the $\taubf_z$ update, whereas in \ALG, the update of $\taubf_s$ uses a more recent version of $\taubf_p$.

\subsection{Recovering GAMP from \ALG}

We now show that the mean-updates of the original sum-product GAMP can be recovered by approximating the mean-updates of \ALG.
For simplicity, we suppress the $t$ index on the variance terms.

At any critical point of Algorithm~\ref{algo:ALG}, we must have $\qbf^t = -\Abf\tran\sbf^t$ and $\zbf^t=\Abf\xbf^t$, as shown in \eqref{eq:qsfix}.
If we substitute these two constraints into the $\vbf$-update objective in \eqref{eq:vmin}, we obtain
\beqa
   \lefteqn{  \|\zbf^t+\taubf_p.\sbf^t -\Abf\vbf\|^2_{\taubf_p}
    + \|\xbf^t+\taubf_r.\qbf^t -\vbf\|^2_{\taubf_r} } \nonumber \\
    &=&  \|\Abf(\xbf^t-\vbf)+\taubf_p.\sbf^t \|^2_{\taubf_p}
    + \|\xbf^t-\vbf-\Diag(\taubf_r)\Abf\tran\sbf^t \|^2_{\taubf_r}. \nonumber
\eeqa
It can be verified that the minimum for this function occurs at $\vbf=\xbf^t$.
So, if we substitute $\vbf^t=\xbf^t$ and $\qbf^t = -\Abf\tran\sbf^t$ into the mean updates in Algorithm~\ref{algo:ALG}, we obtain
\beqan
    \xbf^{\tp1} &=& g_x(\rbf^t,\taubf_r), \\
    \zbf^{\tp1} &=& g_z(\pbf^t,\taubf_p), \\
    \sbf^{\tp1} &=& \sbf^t + \Diag(\one./\taubf_p)(\zbf^{\tp1}-\Abf\xbf^t), \\
    \rbf^{\tp1} &=& \xbf^{\tp1} + \Diag(\taubf_r)\Abf\tran\sbf^{\tp1}, \\
    \pbf^{\tp1} &=& \Abf\xbf^{\tp1} - \taubf_p.\sbf^{\tp1} .
\eeqan
Then, substituting the $\pbf$ update into the $\sbf$ update, defining $\zbfbar^t=\zbf^{\tp1}$ and $\sbfbar^t=\sbf^{\tp1}$, and reordering the steps, we obtain
\beqan
    \pbf^{t} &=& \Abf\xbf^{t} - \taubf_p\sbfbar^{\tm1}, \\
    \zbfbar^{t} &=& g_z(\pbf^t,\taubf_p), \\
    \sbfbar^{t} &=& \Diag(\one./\taubf_p)(\zbfbar^{t}-\pbf^{t}), \\
    \rbf^{t} &=& \xbf^{t} + \Diag(\taubf_r)\Abf\tran\sbfbar^{\tm1}, \\
    \xbf^{\tp1} &=& g_x(\rbf^t,\taubf_r) ,
\eeqan
which is precisely the GAMP mean-update loop.

\section{Numerical Experiments} \label{sec:Simulations}

We now illustrate the performance of ADMM-GAMP by considering three numerical experiments.
While our theoretical results assumed strictly convex penalties, we numerically demonstrate the
stability of \ALG for the non-convex penalty corresponding to a
Bernoulli-Gaussian prior on $\xbf$, i.e.,
\begin{equation}
p_x(x) = (1-\rho) \delta(x) + \rho \mathcal{N}(x; 0, 1),
\end{equation}
where $\rho \in (0, 1]$ is the sparsity ratio and $\delta$ is the Dirac delta distribution.
In our experiments, we fix the parameters to $n = 1000$ and $\rho = 0.2$, and we numerically compare the normalized MSE
\begin{equation*}
\text{NMSE (dB)} \defn 10\log_{10}\left(\frac{\|\xbf - \xbfhat\|_2^2}{\|\xbf\|_2^2}\right)
\end{equation*}
of ADMM-GAMP to four other recovery schemes:
the original GAMP method~\cite{Rangan:11-ISIT};
de-biased LASSO~\cite{Tibshirani:96};
swept AMP (SwAMP)~\cite{manoel2014swamp};
and the support-aware MMSE estimator, labeled ``genie."
The SwAMP method is identical to original GAMP method
but updates only one component of $\xbf$ at a time
-- a common technique also used for stabilizing loopy BP\@.
For LASSO, we optimized the regularization parameter $\lambda$ for best MSE performance.
For GAMP, SwAMP, and ADMM-GAMP, we terminated the iterations as soon as $\|\xbfhat^t - \xbfhat^{t-1}\|_2/\|\xbfhat^{t-1}\|_2 \leq 10^{-4}$
and imposed an upper limit of $200$ iterations.
\textb{In all experiments below, \ALG was run with 10 iterations of the inner loop
ADMM minimization for each outer loop update.  Also, the least-sqaures 
minimization \eqref{eq:vmin} was performed with 3 conjugate gradient
iterations per inner loop iteration, using as a warm start, the output of final value
from the previous iteration as the initial condition of the current iteration.}

\begin{figure}
\begin{center}
\footnotesize
\includegraphics[width=8.5cm]{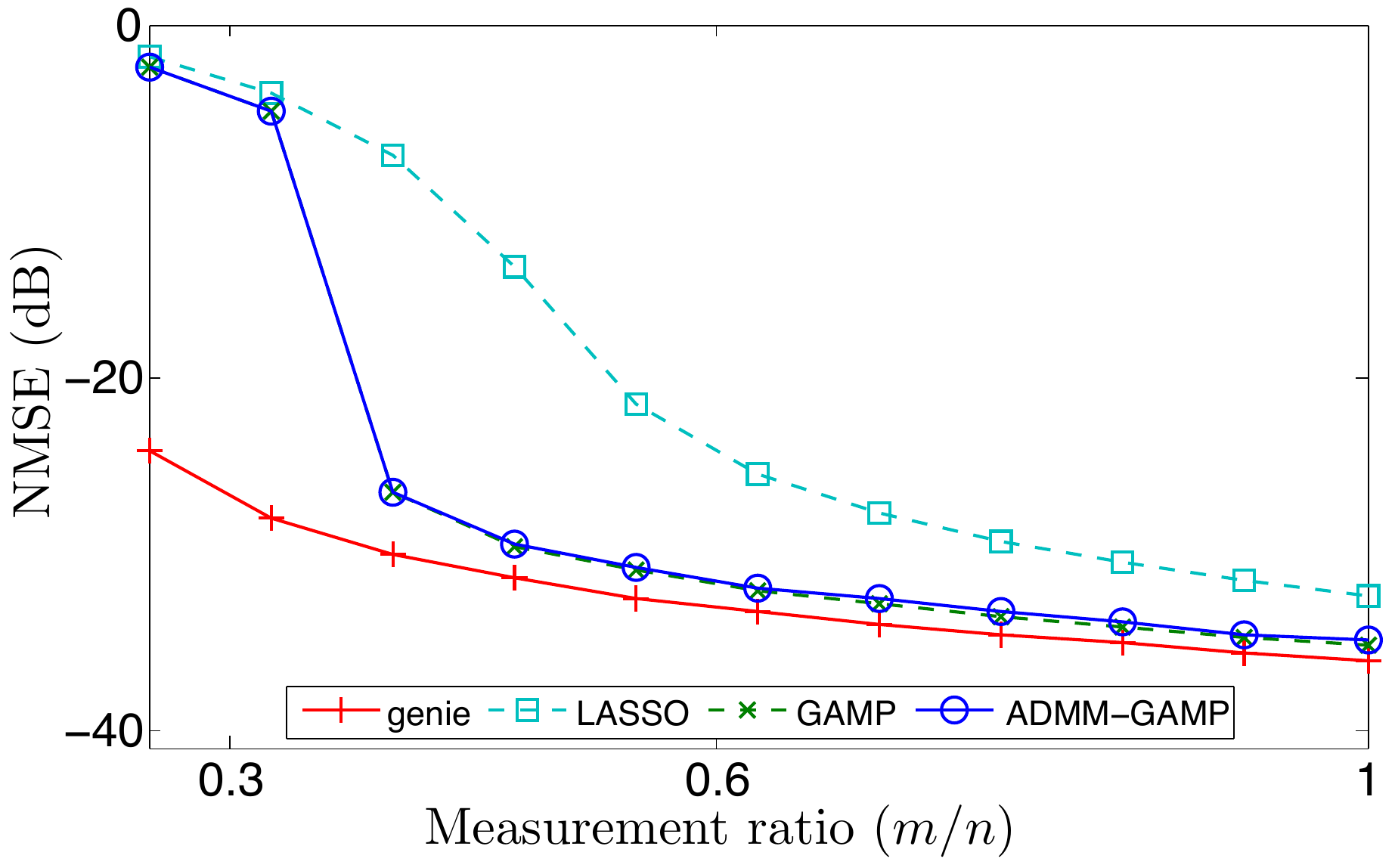}
\end{center}
\caption{Average NMSE
versus measurement rate $m/n$
when recovering a length $n=1000$ Bernoulli-Gaussian signal $\xbf$
from AWGN-corrupted measurements $\ybf=\Abf\xbf+\ebf$
under i.i.d.\ $\Abf$.
}
\label{fig:measurementsVsNmse}
\end{figure}

In our first experiment, we consider a standard problem: recover sparse $\xbf$ from
$\ybf = \Abf\xbf + \ebf$, where $\ebf$ is AWGN with variance set to achieve an SNR of $30$ dB, and where the measurement matrix $\Abf$ is drawn with i.i.d.\ $\mathcal{N}(0, 1/m)$ entries.
Figure~\ref{fig:measurementsVsNmse} shows the NMSE performance of the algorithms under test after averaging the results of $100$ Monte Carlo trials.
Here, since $\ybf$ and $\zbf=\Abf\xbf$ are related through AWGN,
the GAMP algorithm of \cite{Rangan:11-ISIT} reduces
to the Bayesian version of the AMP algorithm from \cite{DonohoMM:10-ITW1}.

Note that the case of i.i.d.\ $\Abf$ is the ``ideal" scenario for both AMP and GAMP\@.   As discussed in the Introduction, their convergence in this case
is guaranteed rigorously through state evolution analysis~\cite{BayatiM:11,Rangan:11-ISIT,JavMon:12-arXiv} as $m,n\rightarrow\infty$.
In Figure~\ref{fig:measurementsVsNmse}, since $m$ and $n$ are sufficiently large, it is not surprising to see that GAMP performs well over all measurement
ratios $m/n$.
Furthermore, it is interesting to notice that GAMP outperforms LASSO and obtains NMSEs that are very close to that of the support-aware genie.
Under such ideal $\Abf$, the proposed \ALG method matches the performance of GAMP
(since it minimizes the same objective) but does
not offer any additional benefit.

The benefits of \ALG become apparent in our second experiment, which uses non-i.i.d.\ matrices $\Abf$.
In describing the experiment, we first recall that
\cite{RanSchFle:14-ISIT} established that the convergence
of GAMP can be predicted by the peak-to-average ratio of the squared
singular values,
\beq \label{eq:kapDef}
    \kappa(\Abf) \defn \frac{\sigma_1^2(\Abf)}{\sum_{i=1}^r \sigma_i^2(\Abf)/r},
\eeq
where $r= \min\{m,n\}$ and $\sigma_i(\Abf)$ is the $i$-th largest
singular value of $\Abf$.
When this ratio $\kappa$ is sufficiently large, the algorithm will diverge.
Thus, to test the robustness of \ALG, we constructed a sequence of matrices
$\Abf$ with varying $\kappa$, as follows.
First, the left and right singular vectors of $\Abf$ were generated by
drawing an $m\times n$ matrix with i.i.d.\ $\mathcal{N}(0, 1/m)$ entries and taking its singular-value decomposition.
Then, the singular values of $\Abf$ were chosen by setting the largest at $\sigma_1(\Abf) = 1$ and logarithmically spacing each successive singular value to attain the desired peak-to-average ratio $\kappa$.

\begin{figure}
\begin{center}
\footnotesize
\includegraphics[width=8.5cm]{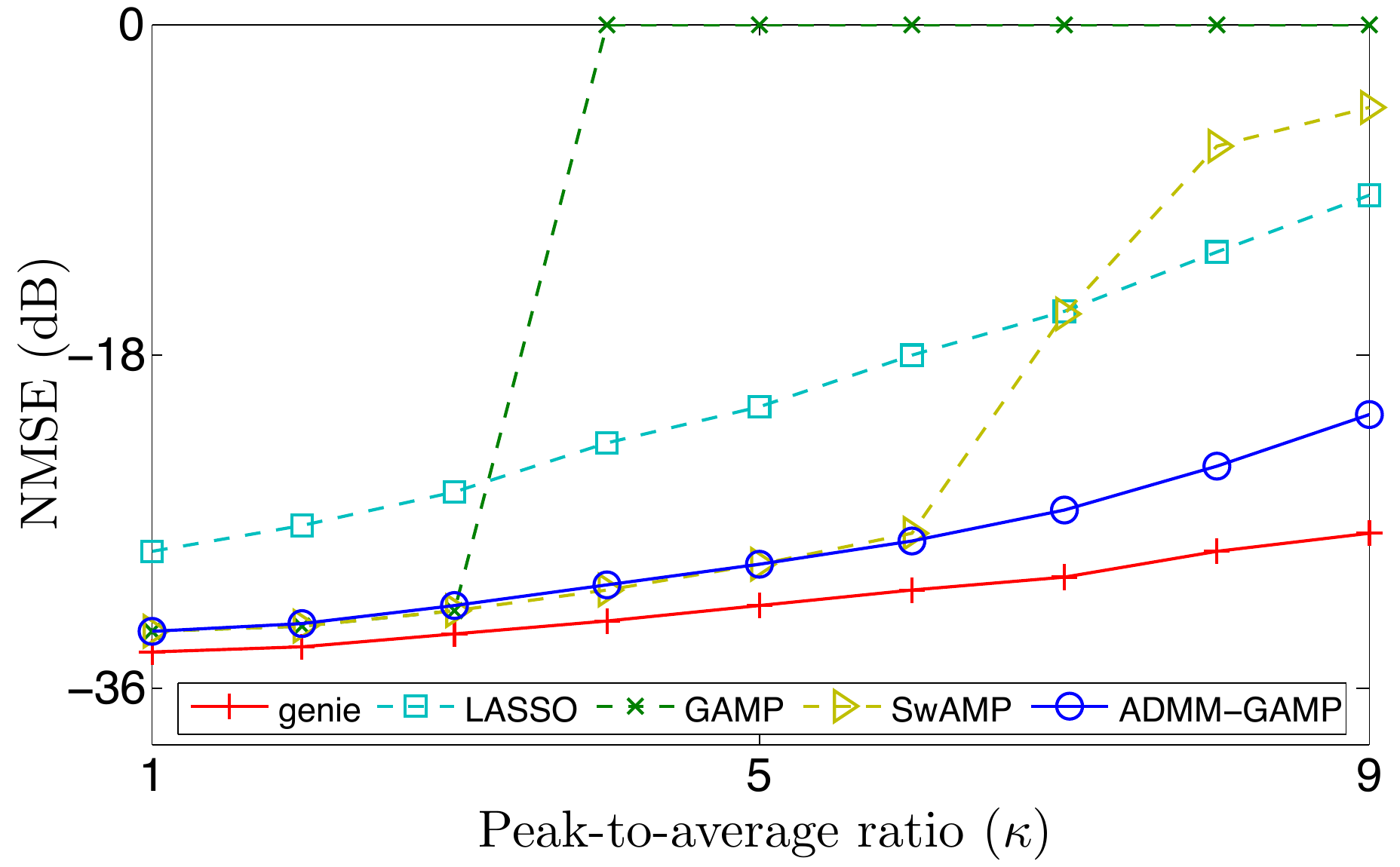}
\end{center}
\caption{Average NMSE
versus peak-to-average squared-singular-value ratio $\kappa(\Abf)$
when recovering a length $n=1000$ Bernoulli-Gaussian signal $\xbf$
from $m=600$ AWGN-corrupted measurements $\ybf=\Abf\xbf+\ebf$.
Note the superior performance of ADMM-GAMP relative to both the original GAMP and SwAMP, and the proximity of ADMM-GAMP to the support-aware genie.}
\label{fig:conditionNumberVsNmse}
\end{figure}

As a function of $\kappa$, the NMSE performance of the various algorithms under test is illustrated in Figure~\ref{fig:conditionNumberVsNmse} for the case of $m = 600$ measurements.
There it can be seen that, for larger values of $\kappa$, the NMSE performance of the original GAMP algorithm deteriorated, which was a result of the algorithm diverging.
(Note that, in the plot, we capped the maximum NMSE to $0$~dB for visual clarity.)
The figure also shows that the SwAMP method achieved low NMSE over a wider range of $\kappa$ ratios than the original GAMP method, but its performance also degraded for larger values of $\kappa$.
The ADMM-GAMP method, however, converged over the entire range of $\kappa$ values, achieving NMSE performance relatively close to the support-aware genie.

\begin{figure}
\begin{center}
\footnotesize
\includegraphics[width=8.5cm]{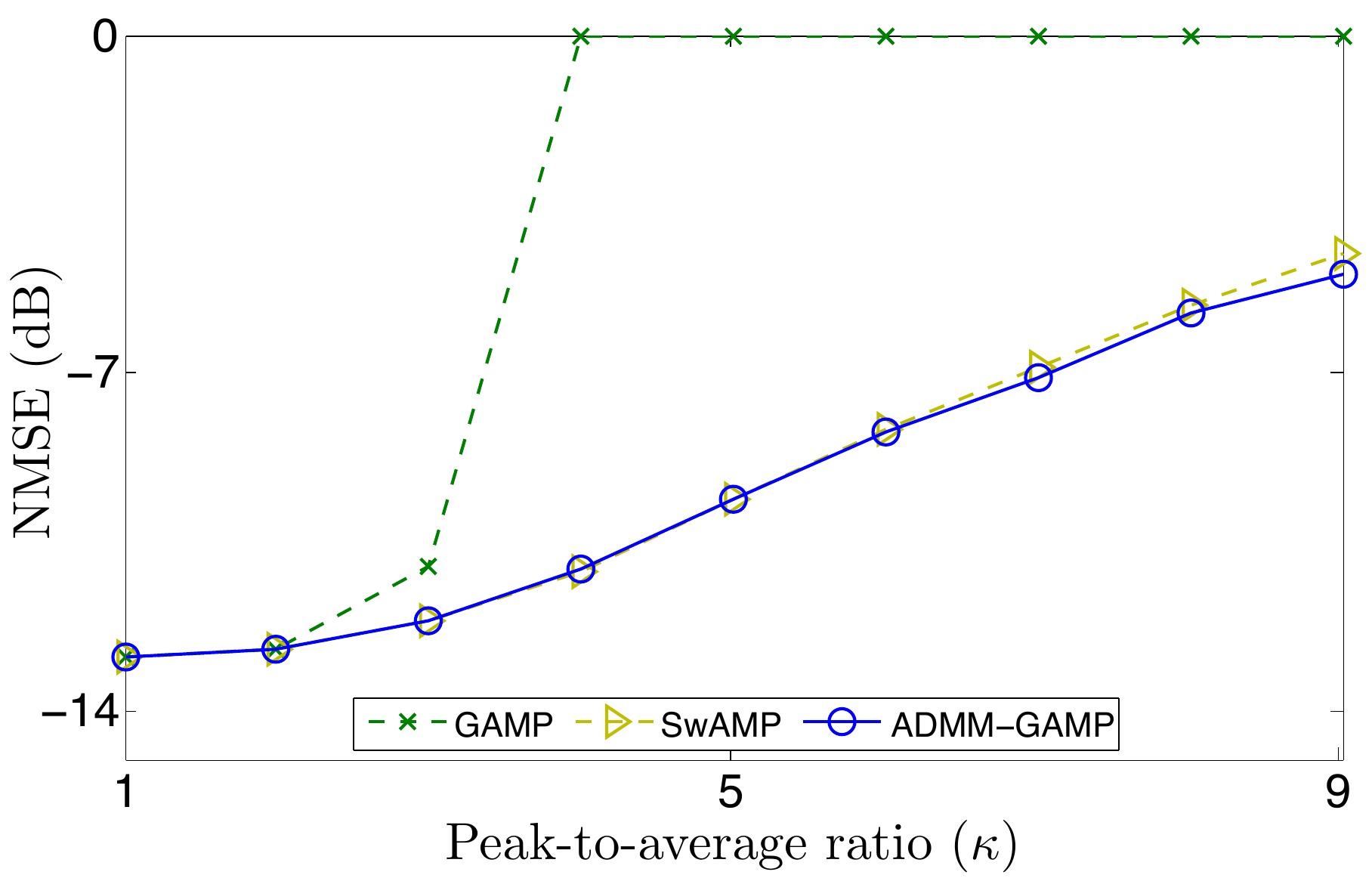}
\end{center}
\caption{Average NMSE
versus peak-to-average squared-singular-value ratio $\kappa(\Abf)$
when recovering a length $n=1000$ Bernoulli-Gaussian signal $\xbf$
from $m=2000$ noiseless 1-bit measurements $\ybf=\mathrm{sgn}(\Abf\xbf)$.
Note the superior performance of ADMM-GAMP relative to the original GAMP and SwAMP.}
\label{fig:OneBitCS}
\end{figure}

In our third and final experiment,
we recover $\xbf$ from ``one-bit" measurements $\ybf = \mathrm{sgn}(\Abf\xbf)$, where $\mathrm{sgn}$ is the \textit{sign function}, as considered in, e.g., \cite{boufounos2008onebit} and \cite{KamilovGR:12}.
Here, we used $m = 2000$ measurements and generated the matrices $\Abf$ as in our second experiment.
Figure~\ref{fig:OneBitCS} shows the NMSE performance of the various algorithms under test.
The results in the figure illustrate that the original GAMP method diverged for $\kappa \geq 2$.
However, both SwAMP and ADMM-GAMP recovered the solution for the whole range of $\kappa$ without diverging, with ADMM-GAMP yielding slightly better NMSE (about $0.3$ dB better) at higher values of $\kappa$.

\section*{Conclusions}

Despite many promising results of AMP methods,
the major stumbling block to more widespread use is their convergence
and numerical stability.  Although AMP techniques admit provable
guarantees for i.i.d.\ $\Abf$, they can easily diverge for transforms
that occur in many practical problems.
While several methods have been proposed
to improve the convergence, this paper provides a method with
provable guarantees under arbitrary transforms.
The method leverages well-established concepts of double-loop methods
in belief propagation \cite{yuille2002concave} as well as the classic
ADMM method in optimization \cite{BoydPCPE:09}.

Nevertheless, there is still much work to be done.
Most obviously, the proposed \ALG method comes at a computational cost.
Each iteration requires solving a (potentially large) least squares problem
\eqref{eq:vmin} that is not needed in the original
AMP and GAMP algorithms.  Similar to standard applications of
ADMM, this minimization can likely be performed via conjugate
gradient iterations, but its implementation requires further study.
In any case, it is possible that \ALG will be slower
than other variants of GAMP\@.
Indeed, our simulations suggest that other methods such as SwAMP
or adaptively damped GAMP \cite{Vila:ICASSP:15} may provide equally robust performance with
less cost per iteration.  One line of future work would thus
be see to whether
the proof techniques in this paper can be extended to address these algorithms
as well.


The analysis in this paper might also be extended to other variants of
AMP and GAMP\@.  For example, it is conceivable that similar analysis
could be applied to develop convergent approaches to the expectation-maximization (EM) GAMP developed in
\cite{KrzMSSZ:11-arxiv,Vila:TSP:13,Vila:TSP:14,KamRanFU:12-nips,KamRanFU:12-IT},
turbo and hybrid GAMP methods in
\cite{SomS:12,RanganFGS:12-ISIT} and applications in
dictionary learning and matrix factorization
\cite{parker2013bilinear,parker2013bilinear2,RanganF:12-ISIT}.

\appendices

\section{Proof of Theorem \ref{thm:iterLinConv}} \label{sec:iterConvPf}
\textb{Throughout this appendix, we use the shorthand notation for the gradient $h'(\taubf)\defn \partial h(\taubf)/\partial \taubf \in \R^p$.}

First we show, by induction, that $\gammabf^k \in \Gamma$ for all $k$.
Recall that, by the hypothesis of the theorem,
$\gammabf^0 \in \Gamma$.  Now suppose that $\gammabf^k \in \Gamma$.
Then the updates in Algorithm~\ref{algo:IterLin} imply that
\[
    h'(\taubf^k) = h'(\gbf(\bbf^k)) = h'(\gbf(\bbfhat(\gammabf^k))).
\]
Then, by Assumption~\ref{as:IterLinConv}(c),
$h'(\taubf^k) \in \Gamma$.  Since $\gammabf^k \in \Gamma$,
$\theta^k \in (0,1]$, and $\Gamma$ is convex,
\[
    \gammabf^{\kp1}
    = (1-\theta^k)\gammabf^k + \theta^k h'(\taubf^k) \in \Gamma.
\]
Thus, by induction, $\gammabf^k \in \Gamma$ for all $k$.

Next, we prove the decrementing property \eqref{eq:JgenDec}.
First observe that
since the restriction $\bbf \in B$ is a linear constraint,
we can find a linear transform $\Bbf$ and vector $\bbf_0$ such that
$\bbf \in B$ if and only if $\bbf=\Bbf\xbf + \bbf_0$ for some vector $\xbf$.
It can be verified that we can reparameterize the functions $f(\cdot)$ and
$\gbf(\cdot)$ around $\xbf$ and obtain the exact same recursions in
Algorithm~\ref{algo:IterLin}.  Also, all the conditions in Assumptions
\ref{as:IterLinConv} will hold for reparametrized functions as well.
Thus, for the remainder of the proof we can ignore the linear constraints
$B$, or alternatively view $B$ as the entire vector space.

Under this assumption, for any $\gammabf$, and any minimizer $\bbfhat(\gammabf)$ will be in the interior of $B$ and therefore,
\beqa
    \zero
    &=& \frac{\partial J(\bbfhat(\gammabf),\gammabf)}{\partial \bbf}
    = f'(\bbfhat(\gammabf))
        + \sum_{\ell=1}^L g_\ell'(\bbfhat(\gammabf))\gamma_\ell
        \label{eq:JderivZero2} \\
    &=& f'(\bbfhat(\gammabf)) + g'(\bbfhat(\gammabf))\gammabf
        \label{eq:JderivZero} ,
\eeqa
where
$f'(\bbf)$ is shorthand notation for the gradient $\partial f(\bbf)/\partial \bbf$,
$g_\ell'(\bbf)$ is shorthand for the gradient (with respect to $\bbf$) of the $\ell$th component of the vector-valued function $\gbf(\cdot)$,
and where $g'(\bbf)=[g_1'(\bbf),\dots,g_L'(\bbf)]$ is matrix-valued.
Taking the gradient of \eqref{eq:JderivZero2} with respect to $\gammabf\tran$ yields the matrix
\beqa
    \zero
    &\stackrel{(a)}{=}&
        \frac{\partial f'(\bbfhat(\gammabf))}{\partial\gammabf\tran}
        + \sum_{\ell=1}^L
        \frac{\partial g_\ell'(\bbfhat(\gammabf))}{\partial\gammabf\tran}
                \gamma_\ell + g'(\bbfhat(\gammabf)) \nonumber\\
    &\stackrel{(b)}{=}&
        \left[ \frac{\partial f'(\bbfhat(\gammabf))}{\partial\bbfhat\tran}
        + \sum_{\ell=1}^L
        \frac{\partial g_\ell'(\bbfhat(\gammabf))}{\partial\bbfhat\tran} \gamma_\ell
        \right] \frac{\partial \bbfhat(\gammabf)}{\partial \gammabf\tran}
        + g'(\bbfhat(\gammabf)) \nonumber \\
    &=& \Hbf(\gammabf) \frac{\partial \bbfhat(\gammabf)}{\partial \gammabf\tran}
        + g'(\bbfhat(\gammabf)) , \label{eq:JhessZero}
\eeqa
where (a) and (b) follow from the chain rule and
$\Hbf(\gammabf)$ is the Hessian from \eqref{eq:Dbnd}.
Equation~\eqref{eq:JhessZero} then implies
\beq \label{eq:bhatDeriv}
    \frac{\partial \bbfhat(\gammabf)}{\partial \gammabf\tran}
    = - \Hbf(\gammabf)^{-1}g'(\bbfhat(\gammabf)),
\eeq
where Assumption~\ref{as:IterLinConv}(b) guarantees the existence of the inverse.
The gradient of the objective with respect to $\gammabf\tran$ is then
\beqa
    \frac{\partial J(\bbfhat(\gammabf))}{\partial \gammabf\tran}
    &\stackrel{(a)}{=}& \left[ f'(\bbfhat(\gammabf))
        + g'(\bbfhat(\gammabf))h'(\taubf)\right]\tran
    \frac{\partial \bbfhat(\gammabf)}{\partial \gammabf\tran}
    \nonumber \\
    &\stackrel{(b)}{=}& \big( h'(\taubf) -\gammabf \big)\tran
    g'(\bbfhat(\gammabf))\tran
    \frac{\partial \bbfhat(\gammabf)}{\partial \gammabf\tran}
        \nonumber \\
    &\stackrel{(c)}{=}& \big( \gammabf -h'(\taubf)\big)\tran
    g'(\bbfhat(\gammabf))\tran\Hbf(\gammabf)^{-1}g'(\bbfhat(\gammabf))
        \label{eq:JderD} , \qquad
\eeqa
where (a) follows from \eqref{eq:JoptGen} and the chain rule,
(b) follows from \eqref{eq:JderivZero}, and
(c) follows from \eqref{eq:bhatDeriv}.

Notice that the $\gammabf^k$ update in Algorithm~\ref{algo:IterLin}
can be written as
\[
    \gammabf^{\kp1}-\gammabf^k = \big(h'(\taubf^k) -\gammabf^k\big)\theta^k.
\]
Taking an inner product of the above and \eqref{eq:JderD} evaluated at $\gammabf=\gammabf^k$, we get
\beqa
    \lefteqn{
    \left[ \frac{\partial J(\bbfhat(\gammabf^k))}{\partial \gammabf\tran} \right]
    \left( \gammabf^{\kp1}-\gammabf^k \right) }\nonumber\\
    &=& - \big( \gammabf^k -h'(\taubf^k)\big)\tran g'(\bbf^k)\tran
        \Hbf(\gammabf^k)^{-1}
        g'(\bbf^k)
        \big( \gammabf^k -h'(\taubf^k)\big) \theta^k \nonumber  \\
    &\leq& -\frac{\theta^k}{c_2} \big\|
        g'(\bbf^k)\big(\gammabf^k -h'(\taubf^k)\big)\big\|^2,
\eeqa
recalling that $\bbfhat(\gamma^k)=\bbf^k$
and that $c_2$ was defined in Assumption~\ref{as:IterLinConv}(b).
Therefore, the update of $\gammabf^k$ is in a descent direction
on the objective $J(\bbfhat(\gammabf))$.  Hence, for a sufficiently
small damping parameter $\theta^k$, we will have
\[
    J(\bbf^{\kp1}) - J(\bbf^k) = J(\bbfhat(\gammabf^{\kp1}))-
    J(\bbfhat(\gammabf^{k})) \leq 0,
\]
which proves the decrementing property \eqref{eq:JgenDec}.

\section{Large Deviations View of MAP Estimation}
\label{sec:largeDev}

For each $T>0$, let $\xbf^t(T), \zbf^t(T), \ldots$, be the output
of the \ALG algorithm with the \SP estimation functions \eqref{eq:GxzSP}
and the scaled penalties \eqref{eq:fxzScale}.
Next, we define several limits.
For the mean vectors we define
\[
    \xbf^t = \lim_{T\arr 0} \xbf^t(T), \quad
    \zbf^t = \lim_{T \arr 0} \zbf^t(T),
\]
for the dual vectors we define
\[
    \qbf^t = \lim_{T \arr 0} T\qbf^t(T), \quad
    \sbf^t = \lim_{T \arr 0} T\sbf^t(T),
\]
and for the variance terms we define
\begin{subequations}\label{eq:tauscale}
\beqa
    \taubf_x^t &=& \lim_{T \arr 0} \frac{\taubf_x^t(T)}{T},
    \taubf_z^t = \lim_{T \arr 0} \frac{\taubf_z^t(T)}{T},
    \taubf_p^t = \lim_{T \arr 0} \frac{\taubf_p^t(T)}{T}, \qquad\\
    \taubf_r^t &=& \lim_{T \arr 0} \frac{\taubf_r^t(T)}{T},
    \taubf_s^t = \lim_{T \arr 0} T\taubf_s^t(T).
\eeqa
\end{subequations}
We will assume that all of these limits exist.
Note that some of terms are scaled by $T$ and others by $1/T$.
These normalizations are important.
It is easily checked that the scalings all cancel, so that the limiting
values satisfy the recursions of Algorithm~\ref{algo:ALG} with the
limiting estimation functions
\begin{subequations} \label{eq:gxLimT}
\beqa
    g_x(\rbf,\taubf_r) &\defn& \lim_{T \arr 0} g_x(\rbf,\taubf_r(T);T)
        = \lim_{T \arr 0} g_x(\rbf,\taubf_r T;T), \qquad \\
    g_z(\pbf,\taubf_p) &\defn& \lim_{T \arr 0} g_z(\pbf,\taubf_p(T);T)
        = \lim_{T \arr 0} g_z(\pbf,\taubf_p T;T),
\eeqa
\end{subequations}
where $g_x(\rbf,\taubf_r T;T)$ and
$g_z(\pbf,\taubf_p T;T)$  are the \SP estimation functions
\eqref{eq:GxzSP} for the
scaled penalties \eqref{eq:fxzScale}.
Note that we have used the scalings in \eqref{eq:tauscale},
which show $\taubf_r(T) \approx T\taubf_r$ and $\taubf_p(T) \approx
\taubf_p T$ for small $T$.
Now, the scaled function $g_x(\rbf,\taubf_r T;T)$ is the
expectation $\Exp(\xbf|T)$ with respect to the density
\[
    p(\xbf|\rbf,\taubf_rT;T) \propto \exp\left[
        -\frac{\fx(\xbf)}{T} - \frac{1}{2T}\|\xbf-\rbf\|^2_{\taubf_r} \right].
\]
Laplace's Principle \cite{DemboZ:98} from large deviations theory
shows that (under mild conditions) this density
concentrates around its maxima, and thus the expectation with
respect to this density converges to the minimum
\[
    \lim_{T \arr 0} g_x(\rbf,\taubf_rT;T) =
     \argmin_{\xbf} \fx(\xbf) + \tfrac{1}{2}
    \|\xbf-\rbf\|^2_{\taubf_r},
\]
which is exactly the minimization in the \MS estimation
function \eqref{eq:GxzMS}.  The limit of $g_z(\pbf,\taubf_p/T;T)$ as
$T\rightarrow 0$ is similar.  We conclude that the limit of the \ALG
algorithm with \SP estimation functions \eqref{eq:GxzSP} and
scaled densities \eqref{eq:fxzScale} is exactly the
\ALG algorithm with the \MS estimation functions \eqref{eq:GxzMS}.
In particular, for each $T$, the density over $x_j$ in \eqref{eq:bxz}
is given by
\beq \label{eq:bxjT}
    b_{x_j}^t(x_j|r_j,\tau_{r_j}T) \propto
    \exp\left[ - \frac{\fxj(x_j)}{T} - \frac{(x_j-r_j^t)^2}{2T\tau_{r_j}^t}
    \right],
\eeq
from which we can prove the limits in \eqref{eq:phihat}.

It remains to show that the LSL-BFE in \eqref{eq:Fbethe}
with the scaled functions \eqref{eq:fxzScale} decomposes
into the optimizations \eqref{eq:xzmap} and \eqref{eq:JMAP2Opt}
as $T \arr 0$.  To this end, let $J(b_x,b_z;T)$ be the LSL-BFE
\eqref{eq:Fbethe} for the scaled penalties \eqref{eq:fxzScale},
which is given by
\beqa
     J(b_x,b_z;T)
     &=&  D(b_x\| Z_{x,T}^{-1}\e^{-\fx/T}) + D(b_z\|Z_{z,T}^{-1}\e^{-\fz/T})
        \nonumber\\&&\mbox{}
     	+ H(\var(\xbf|b_x),\var(\zbf|b_z))  \nonumber \\
     &=& \frac{1}{T}\Big[ \Exp( \fx(\xbf)|b_x) + \Exp(\fz(\zbf)|b_z) \Big]
     \nonumber \\
	&& \mbox{} + H(\var(\xbf|b_x),\var(\zbf|b_z)) \nonumber \\
    && \mbox{} - H(b_x) - H(b_z)
	+ \text{const}, \label{eq:JBFET}
\eeqa
where $H(a)$ denotes the differential entropy of distribution $a$,
$H(\taubf_x,\taubf_z)$ is the entropy bound from \eqref{eq:Hgauss},
$\ZxT\defn\int \e^{-\fx(\xbf)/T} \d\xbf$,
$\ZzT\defn\int \e^{-\fz(\zbf)/T} \d\zbf$,
and the ``const" in \eqref{eq:JBFET} is with respect to $b_x$ and $b_z$.
Now, we know that, as $T \arr 0$, the optimal densities $b_x$ and $b_z$
will concentrate around their maxima with variance $O(T)$.
Thus, we can take a quadratic approximation around the maximum
\beq
    \ln b_{x_j}(x_j) \approx \frac{(x_j-\xhat_j)^2}{2T\tau_{x_j}} + \text{const},
\eeq
where
\[
    \xhat_j = \argmin_{x_j} -\ln b_{x_j}(x_j), \quad
    \frac{1}{\tau_{x_j}} = -
        T\frac{\partial^2 \ln b_{x_j}(x_j)}{\partial x_j^2},
\]
with a similar approximation for $\ln b_{z_i}(z_i)$.
Under these approximations, $b_{x_j}(x_j)$
and $b_{z_i}(z_i)$ become approximately Gaussian, i.e.,
\beq \label{eq:bxzN}
    b_{x_j}(x_j) \approx {\mathcal N}(\xhat_j,T\tau_{x_j}),
    \quad
    b_{z_i}(z_i) \approx {\mathcal N}(\zhat_i,T\tau_{z_i}).
\eeq
Using these Gaussian approximations, we can compute the expectations
\beqa
    \lefteqn{
    \Exp(\fxj(x_j)|b_{x_j})
    = \int \fxj(x) \mathcal{N}(x;\xhat_j,T\tau_{x_j}) \d x
    }\nonumber\\
    &\stackrel{(a)}{=}& \int \sum_{k=0}^\infty
        \frac{(x-\xhat_j)^k \fxj^{(k)}(\xhat_j)}{k!}
	\mathcal{N}(x;\xhat_j,T\tau_{x_j}) \d x \\
    &\stackrel{(b)}{=}& \sum_{k=0}^\infty
        \frac{\fxj^{(k)}(\xhat_j)}{k!}
	\int (x-\xhat_j)^k \mathcal{N}(x;\xhat_j,T\tau_{x_j}) \d x \\
    &\stackrel{(c)}{=}& \sum_{l=0}^\infty
        \frac{\fxj^{(2l)}(\xhat_j)}{(2l)!}
	(T\tau_{x_j})^l (2l-1)!! \\
    &\stackrel{(c)}{=}& \sum_{l=0}^\infty
        \frac{\fxj^{(2l)}(\xhat_j)}{2^l l!}
	(T\tau_{x_j})^l ,
\eeqa
where (a) wrote $\fxj(x)$ using a Taylor series about $x=\xhat_j$;
(b) assumed the exchange of limit and integral;
(c) used the expression for the Gaussian central moments, which involves the double factorial $(2l-1)!!=(2l-1)(2l-3)(2l-5)\times\dots\times 1$;
and (d) used the identity $(2l-1)!!=\frac{(2l)!}{2^l l!}$.
Thus, for small $T$, we have
\begin{subequations} \label{eq:fxzT}
\beqa
    \Exp(\fxj(x_j)|b_{x_j})
    &\approx & \fxj(\xhat_j)
    	+ \frac{1}{2} T\tau_{x_j}\fxj''(\xhat_j), \\
    \Exp(\fzi(z_i)|b_{z_i})
    &\approx &  \fzi(\zhat_i)
    	+ \frac{1}{2} T\tau_{z_i}\fzi''(\zhat_i).
\eeqa
\end{subequations}

The differential entropies of these Gaussians \eqref{eq:bxzN}
are
\beq \label{eq:HbxzT}
    H(b_{x_j}) = \frac{1}{2}\ln(2\pi e T\tau_{x_j}), \quad
    H(b_{z_i}) = \frac{1}{2}\ln(2\pi e T\tau_{z_i}),
\eeq
and the entropy term \eqref{eq:Hgauss} is
\beqa
    \lefteqn{
    H(\var(\xbf|b_x),\var(\zbf|b_z)) = H(T\taubf_x,T\taubf_z) }
    \nonumber \\
    &=& \frac{1}{2}\left[
    \sum_{i=1}^m \frac{\tau_{z_i}}{\tau_{p_i}} + \ln(2\pi T\tau_{p_i}) \right],
    ~
    \taubf_p \defn \Sbf\taubf_x.
    \label{eq:HguassT}
\eeqa
Substituting \eqref{eq:fxzT}, \eqref{eq:HbxzT} and \eqref{eq:HguassT}
into \eqref{eq:JBFET}, we obtain
\beq \label{eq:JBFE0}
    J_T(b_x,b_z)
    = \frac{1}{T}J(\xbfhat,\zbfhat)
    	+ \frac{1}{2} J^2(\taubf_x,\taubf_z,\xbfhat,\zbfhat) + \text{const},
\eeq
where $J(\cdot)$ and $J^2(\cdot)$ are given in \eqref{eq:xzmap} and \eqref{eq:JMAP2}.
As $T \arr 0$, the first term in \eqref{eq:JBFE0} dominates, implying that the optimization of $(\xbfhat,\zbfhat)$ can be conducted independently of $\taubf_x,\taubf_z$, as in \eqref{eq:xzmap}.
The subsequent optimization of $(\taubf_x,\taubf_z)$ then follows, as given in \eqref{eq:JMAP2Opt}.

\section{Proof of Theorems \ref{thm:fix} and \ref{thm:fixMS} }
\label{sec:fixPf}

We will just prove Theorem~\ref{thm:fix} since the proof of
Theorem~\ref{thm:fixMS} is very similar.
For the original constrained optimization \eqref{eq:FbetheOpt},
define the Lagrangian
\beq \label{eq:LADMM0}
    L_0(b_x,b_z,\sbf) \defn
        J(b_x,b_z) + \sbf\tran(\Exp(\zbf|b_z) - \Abf\Exp(\xbf|b_x)).
\eeq
We need to show that any fixed points $(b_x,b_z,\sbf)$
of ADMM-GAMP are critical points of this Lagrangian.

First observe that, any fixed point, $\taubf_r$
from line \ref{line:taurstep} of Algorithm~\ref{algo:ALG} satisfies
\beq   \label{eq:taurfixDeriv}
    \one./(2\taubf_r) = \one./(2\taubfbar_r)
    = \frac{\partial H(\taubf_x,\taubf_z)}{\partial \taubf_x},
\eeq
where the last step follows from the construction
of $\taubfbar_r$ in \eqref{eq:taubardef}.
Similarly, at any fixed point of line \ref{line:taupstep},
\beq
    \one./(2\taubf_p) = \one./(2\taubfbar_p)
    = \frac{\partial H(\taubf_x,\taubf_z)}{\partial \taubf_z}.
\eeq

From \eqref{eq:ADMMs} and \eqref{eq:ADMMq}, we see that any fixed point
satisfies
\beq \label{eq:zxcon}
    \Exp(\zbf|b_z) = \Abf\vbf, \quad \Exp(\xbf|b_x) = \vbf.
\eeq
Thus, the constraint in \eqref{eq:FbetheOpt} is satisfied, in that
$\Exp(\zbf|b_z) = \Abf\Exp(\xbf|b_x)$.
Furthermore, since $\vbf$ minimizes \eqref{eq:vmin}, we know that it
zeros the gradient of the corresponding cost function:
\beq
   \zero
   = \Abf\tran \Dbf_{\taubf_p}\big(\Exp(\zbf|b_z)-\Abf\vbf+\taubf_p.\sbf\big)
        +\Dbf_{\taubf_r}\big(\Exp(\xbf|b_x)-\vbf+\taubf_r.\qbf\big) ,
\eeq
where $\Dbf_{\taubf}=\Diag(\one./\taubf)$.
Plugging \eqref{eq:zxcon} into the previous expression, we obtain
\beq \label{eq:qsfix}
    \qbf = -\Abf\tran \sbf.
\eeq

Since $b_x$ minimizes the augmented Lagrangian in \eqref{eq:ADMMb},
it zeros the corresponding gradient, i.e.,
\beqa
    \zero &=& \frac{\partial}{\partial b_x} L(b_x,b_z,\sbf,\qbf,\vbf;\taubf_r,\taubf_p) \nonumber \\
    &\stackrel{(a)}{=}& \frac{\partial}{\partial b_x} \Bigl[ J(b_x,b_z,\taubf_r,\taubf_p)
        + \qbf\tran\Exp(\xbf|b_x)
        \nonumber \\ && \mbox{}
        + \frac{1}{2}\|\Exp(\xbf|b_x)-\vbf\|^2_{\taubf_r} \Bigr]
        \nonumber \\
    &\stackrel{(b)}{=}& \frac{\partial}{\partial b_x} \left[ J(b_x,b_z,\taubf_r,\taubf_p)
        - \qbf\tran\Exp(\xbf|b_x) \right] \nonumber \\
    &\stackrel{(c)}{=}& \frac{\partial}{\partial b_x} \left[ J(b_x,b_z,\taubf_r,\taubf_p)
        - \sbf\tran\Abf\Exp(\xbf|b_x) \right] \nonumber \\
    &\stackrel{(d)}{=}& \frac{\partial}{\partial b_x} \Bigg[
        J(b_x,b_z) - H\big(\var(\xbf|b_x),\taubf_z\big)
        \nonumber \\ && \mbox{}
        + \left(\one./(2\taubf_r)\right)\tran\var(\xbf|b_x) - \sbf\tran\Abf\Exp(\xbf|b_x) \Bigg] \nonumber \\
    &\stackrel{(e)}{=}& \frac{\partial}{\partial b_x} \left[
        J(b_x,b_z) - \sbf\tran\Abf\Exp(\xbf|b_x) \right] \nonumber \\
   &\stackrel{(f)}{=}& \frac{\partial}{\partial b_x} L_0(b_x,b_z,\sbf),
   \label{eq:bxmin0}
\eeqa
where (a) follows from substituting \eqref{eq:LaugSP0} and eliminating
terms that do not depend on $b_x$, since their gradient equals zero;
(b) follows from \eqref{eq:zxcon};
(c) follows from \eqref{eq:qsfix};
(d) follows from the definitions of the original and linearized LSL-BFEs in
\eqref{eq:Fbethe} and \eqref{eq:JlinBFE};
(e) follows from the chain rule and the gradient in \eqref{eq:taurfixDeriv};
and (f) follows from \eqref{eq:LADMM0}.
A similar computation shows that
\beq \label{eq:bzmin0}
    \frac{\partial}{\partial b_z} L_0(b_x,b_z,\sbf) = \zero.
\eeq
Together, \eqref{eq:bxmin0} and \eqref{eq:bzmin0} show that
$(b_x,b_z)$ are critical points of the Lagrangian $L_0(b_x,b_z,\sbf)$
for the dual parameters $\sbf$.  Since these densities also
satisfy the constraint $\Exp(\zbf|b_z) = \Abf\Exp(\xbf|b_x)$,
we conclude that $(b_x,b_z)$ are critical points of the
constrained optimization \eqref{eq:FbetheOpt}.

\section{Proof of Lemma \ref{lem:gspConv}} \label{sec:gspConvPf}

For the \MS estimation functions \eqref{eq:GxzMS}, we know that
\[
    \xhat_j = g_{x_j}(r_j,\tau_{r_j})
    = \argmin_{x_j} \fxj(x_j) + \frac{1}{2\tau_{r_j}}(x_j-r_j)^2,
\]
which implies that $x_j=\xhat_j$ is a solution to
$0=\fxj'(x_j)+(x_j-r_j)/\tau_{r_j}$, i.e., that
\[
    \xhat_j
    = r_j - \tau_{r_j} \fxj'(\xhat_j).
\]
Taking the derivative with respect to $r_j$, we find
\[
    \frac{\partial \xhat_j}{\partial r_j}
    = 1 - \tau_{r_j} \fxj''(\xhat_j) \frac{\partial \xhat_j}{\partial r_j},
\]
which can be rearranged to form
\beq \label{eq:gxderivMS}
    \frac{\partial \xhat_j}{\partial r_j} = g_{x_j}'(r_j,\tau_{r_j})
        = \frac{1}{1 + \fxj''(\xhat_j)\tau_{r_j}}.
\eeq
Then, given the assumption in the lemma, \eqref{eq:gxderivMS} implies that
\[
    \frac{1}{1+B\tau_{r_j}} \leq g_{x_j}'(r_j,\tau_{r_j})
    \leq \frac{1}{1+A\tau_{r_j}}.
\]
A similar bound can be obtained for $g_{z_i}'(p_i,\tau_{p_i})$,
which proves \eqref{eq:gxzdbnd} for any fixed $\taubf_r$
and $\taubf_p$.

The proof for \SP estimation functions \eqref{eq:GxzSP}
uses a classic result of log-concave functions \cite{brascamp2002extensions}.
Since the functions $\fx$ and $\fz$ are separable, so are the
estimation functions $g_x$ and $g_z$ \eqref{eq:GxzSP},
as established in \eqref{eq:gxSP}.
In particular, we can write
\[
    g_{x_j}(r_j,\tau_{r_j}) = \Exp(x_j|r_j,\tau_{r_j}), \quad
    g_{z_i}(p_i,\tau_{p_i}) = \Exp(z_i|p_i,\tau_{p_i}),
\]
where the expectations are with respect to the densities
\begin{subequations}
\beqa
    p(x_j|r_j,\tau_{r_j}) &\propto& \exp\left[ -\fxj(x_j)-
        \frac{(x_j-r_j)^2}{2\tau_{r_j}} \right], \label{eq:pxjpf} \\
    p(z_i|p_i,\tau_{p_i}) &\propto& \exp\left[ -\fzi(z_i)-
        \frac{(z_i-p_i)^2}{2\tau_{p_i}} \right].
\eeqa
\end{subequations}
We then need to show that the
condition \eqref{eq:gxzdbnd} is satisfied for each of the functions
$g_{x_j}$ and $g_{z_i}$.
Below, we prove this for $g_{x_j}$, noting that
the proof for $g_{z_i}$ is similar.

From \eqref{eq:tauxSP}, we know that
the derivative of $g_{x_j}(r_j,\tau_{r_j})$ with respect to $r_j$ is given by
\beq \label{eq:gxjderiv}
    g_{x_j}'(r_j,\tau_{r_j}) =
        \frac{\tau_{x_j}}{\tau_{r_j}}, \quad
        \tau_{x_j} = \var(x_j|r_j,\tau_{r_j}).
\eeq
The variance here is with respect to the
density \eqref{eq:pxjpf}, which can be rewritten as
\[
    p(x_j|r_j,\tau_{r_j}) = \exp\left[ -h(x_j) \right]
\]
for the potential function
\[
    h(x_j) = \fxj(x_j) +
        \frac{(x_j-r_j)^2}{2\tau_{r_j}},
\]
which has second derivative
\[
    h''(x_j) = \fxj''(x_j) + \frac{1}{\tau_{r_j}}.
\]
By assumption \eqref{eq:fxzderivBnd}, this derivative
is bounded as
\[
    A + \frac{1}{\tau_{r_j}} \leq h''(x_j)
    \leq B + \frac{1}{\tau_{r_j}}.
\]
In particular, $h(x_j)$ is strictly convex.
From \eqref{eq:gxjderiv} and
\cite[Theorem 4.1]{brascamp2002extensions}, we have that
\beqa
    \lefteqn{ g_{x_j}'(r_j,\tau_{r_j}) = \frac{\tau_{x_j}}{\tau_{r_j}} =
    \frac{\var(x_j|r_j,\tau_{r_j})}{\tau_{r_j}} } \nonumber \\
    && \leq \frac{1}{\tau_{r_j}}\Exp\left( \frac{1}{h''(x)} \right) \leq
        \frac{1}{A\tau_{r_j} + 1}. \label{eq:gxderUp}
\eeqa
It is also shown in equation (4.13) of
\cite{brascamp2002extensions} that
\beqa
    \lefteqn{ g_{x_j}'(r_j,\tau_{r_j}) = \frac{\tau_{x_j}}{\tau_{r_j}} =
    \frac{\var(x_j|r_j,\tau_{r_j})}{\tau_{r_j}} } \nonumber \\
    && \geq \frac{1}{\Exp(h''(x_j))\tau_{r_j}} \geq
        \frac{1}{B\tau_{r_j} + 1}. \label{eq:gxderLo}
\eeqa
Thus, we conclude that
\[
    \frac{1}{1+B\tau_{r_j}} \leq g_{x_j}'(r_j,\tau_{r_j})
    \leq \frac{1}{1+A\tau_{r_j}},
\]
which proves \eqref{eq:gxzdbnd}.

\section{Proof of Theorem \ref{thm:meanConv}} \label{sec:meanConvPf}

We find it easier to analyze the algorithm after the variables are combined and
scaled as
\beq \label{eq:tauBlk}
    \taubf \defn \left[ \begin{array}{c} \taubf_r \\ \taubf_p \end{array} \right],
    \quad
    \Dbf \defn \Diag(\one./\taubf),
\eeq
and
\beq \label{eq:blk}
    \wbf \defn \Dbf^{1/2}\left[ \begin{array}{c} \xbf \\ \zbf \end{array} \right], ~
    \ubf \defn \Dbf^{-1/2}\left[ \begin{array}{c} \qbf \\ \sbf \end{array} \right], ~
    \Bbf \defn \Dbf^{1/2}\left[ \begin{array}{c} \Ibf \\ \Abf \end{array} \right].
\eeq
Also, we define
\beq \label{eq:gdefw}
    g(\wbf,\taubf) \defn \left[ \begin{array}{c}
        g_x(\xbf,\taubf_r)\\g_z(\zbf,\taubf_p) \end{array} \right],
\eeq
and henceforth suppress the dependence on $\taubf$ in the notation since $\taubf$  is constant in this analysis.
The mean update steps in Algorithm~\ref{algo:ALG} then become
\begin{subequations}
\beqa
    \wbf^{\tp1}
    &=& \Dbf^{1/2}g(\Dbf^{-1/2}(\Bbf\vbf^{t} - \ubf^t))
    	\label{eq:wtblk} \\
    \ubf^{\tp1}
    &=& \ubf^t + \wbf^{\tp1} - \Bbf\vbf^t
    	\label{eq:utblk} \\
    \vbf^{\tp1}
    &=& \argmin_\vbf \| \wbf^{\tp1} + \ubf^{\tp1} - \Bbf\vbf \|^2,
    	\label{eq:vminblk}
\eeqa
\end{subequations}
where the result of \eqref{eq:vminblk} can be written explicitly as
\beq \label{eq:vminblk2}
    \vbf^t
    = (\Bbf\tran\Bbf)^{-1}\Bbf\tran(\wbf^t + \ubf^t).
\eeq

Let us define
\beq \label{eq:Pperp}
    \Pbf \defn \Bbf(\Bbf\tran\Bbf)^{-1}\Bbf\tran,
    \quad
    \Pbf^\perp \defn \Ibf - \Pbf,
\eeq
where $\Pbf$ is an orthogonal projector operator onto the column space of $\Bbf$
and $\Pbf^\perp$ is the projection onto its orthogonal complement.
Noting that $\Bbf\vbf^t = \Pbf(\wbf^{t} + \ubf^{t})$, \eqref{eq:wtblk} reduces to
\beqa
   \wbf^{\tp1}
   &=& \Dbf^{1/2}g\left(\Dbf^{-1/2}(\Pbf\big(\wbf^t+\ubf^t) - \ubf^t\big)\right) \nonumber\\
   &=& \Dbf^{1/2}g\left(\Dbf^{-1/2}(\Pbf\wbf^t - \Pbf^\perp\ubf^t)\right) \nonumber \\
   &=& \gtilde(\Pbf\wbf^t - \Pbf^\perp\ubf^t),
    \label{eq:wtblk2}
\eeqa
where
\beq \label{eq:gtilde}
    \gtilde(\wbf) \defn \Dbf^{1/2}g(\Dbf^{-1/2}\wbf).
\eeq
Also, since $\Pbf^\perp\Bbf = \zero$, \eqref{eq:utblk} implies that
\beqa
    \Pbf^\perp\ubf^{\tp1}
    &=& \Pbf^\perp\ubf^t + \Pbf^\perp\wbf^{\tp1} \nonumber \\
    &=& \Pbf^\perp\ubf^t + \Pbf^\perp \gtilde(\Pbf\wbf^t - \Pbf^\perp\ubf^t)
    \label{eq:utblk2} .
\eeqa
Now define the state vector
\beq
    \thetabf^t
    \defn \left[ \begin{array}{c}
    	\Pbf \wbf^t \\
	    \Pbf^\perp\ubf^t
	\end{array} \right]
    \label{eq:theta}.
\eeq
Since $\Pbf^2=\Pbf$ and $(\Pbf^\perp)^2 =\Pbf^\perp$,
\[
    \left[ \begin{array}{cc} \Pbf & -\Pbf^\perp \end{array} \right] \thetabf^t
= \Pbf\wbf^t - \Pbf^\perp\ubf^t.
\]
Therefore, from \eqref{eq:wtblk2} and \eqref{eq:utblk2}, respectively, we have that
\beqa
    \Pbf\wbf^{\tp1}
    &=& \Pbf \gtilde\big(
	        \left[ \begin{array}{cc} \Pbf
                & -\Pbf^\perp \end{array} \right] \thetabf^t
	\big), \label{eq:wtblk3} \\
    \Pbf^\perp\ubf^{\tp1}
    &=& \Pbf^\perp\ubf^t + \Pbf^\perp \gtilde\big(
	        \left[ \begin{array}{cc} \Pbf
                & -\Pbf^\perp \end{array} \right] \thetabf^t
        \big) \label{eq:utblk3} .
\eeqa
From \eqref{eq:theta}, \eqref{eq:wtblk3}, and \eqref{eq:utblk3}, we see that
the mean update steps in Algorithm~\ref{algo:ALG} are characterized by
the recursive system
\beq
    \thetabf^{\tp1} = f(\thetabf^t)
    \label{eq:projSys}
\eeq
for
\beq
    f(\thetabf)
    \defn \left[ \begin{array}{c}
	\Pbf \\ \Pbf^\perp
	\end{array} \right]
	\gtilde \big(
	\left[ \begin{array}{cc} \Pbf
                & -\Pbf^\perp \end{array} \right] \thetabf
	\big) + \left[ \begin{array}{cc}
	\zero & \zero \\ \zero & \Pbf^\perp
	\end{array} \right] \thetabf.
    \label{eq:f}
\eeq

The following is a standard contraction mapping result \cite{Vidyasagar:78}:
if $f$ has a continuous Jacobian $f'$
whose spectral norm is less than one, i.e., $\exists \epsilon>0 \text{~s.t.~} \|f'(\thetabf)\|< 1-\epsilon~\forall \thetabf$, then the system \eqref{eq:projSys} converges to a unique fixed point, $\thetabf^*$,
with a \emph{linear} convergence rate, i.e.,
\[
    \exists C>0 \text{~s.t.~}\|\thetabf^t - \thetabf^*\| \leq C(1-\epsilon)^t.
\]
So, our proof will be complete if we can show that
the Jacobian of $f$ from \eqref{eq:f} is indeed a contraction.

First observe that, from the definition of $g(\wbf)$ in \eqref{eq:gdefw},
and the separability and boundedness
assumptions in Assumption~\ref{as:cont}, the Jacobian
of $g(\wbf)$ at any $\wbf$ is diagonal and bounded:
\[
    \exists \epsilon\in(0,0.5] \text{~s.t.~} g'(\wbf) = \Diag( \dbf )\text{~and~}\epsilon \leq d_k \leq 1-\epsilon~\forall k .
\]
Since $\Dbf = \Diag(\one./\taubf)$ is also diagonal, the Jacobian of $\gtilde(\wbf)$
in \eqref{eq:gtilde} is given by
\[
    \gtilde'(\wbf) = \Dbf^{-1/2}\Diag( \dbf )\Dbf^{1/2} = \Diag( \dbf ),
\]
and hence
\beq \label{eq:gtcontract}
    \epsilon \Ibf \leq \gtilde'(\wbf) \leq (1-\epsilon) \Ibf
\eeq
for all $\wbf$.
Now, the Jacobian of $f(\thetabf)$ in \eqref{eq:f} is given by
\beq \label{eq:fderiv1}
    f'(\thetabf) =
    \left[ \begin{array}{c}
	\Pbf \\ \Pbf^\perp
	\end{array} \right]
	\gtilde'(\wbf)
	\left[ \begin{array}{cc} \Pbf & -\Pbf^\perp \end{array} \right]
    + \left[ \begin{array}{cc}
	\zero & \zero \\ \zero & \Pbf^\perp
	\end{array} \right] .
\eeq
Hence, if we define
\beq \label{eq:Jf}
    \Jbf(\thetabf) \defn f'(\thetabf)
    \left[ \begin{array}{cc} \Ibf & \zero \\ \zero & -\Ibf \end{array} \right],
\eeq
then $\|f'(\thetabf)\|=\|\Jbf(\thetabf)\|$ so $f'(\thetabf)$ is a contraction if and
only if $\Jbf(\thetabf)$ is.  Therefore, it suffices to prove that
$\Jbf(\thetabf)$ is a contraction.
Combining \eqref{eq:fderiv1} and \eqref{eq:Jf}, we obtain
\beq \label{eq:fderiv}
    \Jbf(\thetabf) = \Ubf\tran\gtilde'(\wbf)\Ubf - \left[ \begin{array}{cc}
	\zero & \zero \\ \zero & \Pbf^\perp
	\end{array} \right],
\eeq
where $\wbf = \left[ \begin{array}{cc} \Pbf & \Pbf^\perp \end{array} \right] \thetabf$,
and
\beq \label{eq:Udef}
    \Ubf = \left[ \begin{array}{cc} \Pbf & \Pbf^\perp \end{array} \right].
\eeq
Since $\Pbf$ is an orthogonal projection and $\Pbf^\perp$ is the projection onto
the orthogonal complement, $\Ubf$ is an isometry.  That is,
\beq \label{eq:Uonto}
    \Ubf\Ubf\tran = \Pbf + \Pbf^\perp = \Ibf,
\eeq
and hence $\Ubf\tran\Ubf \leq \Ibf$.
Therefore, from \eqref{eq:fderiv} and \eqref{eq:gtcontract},
\beq \label{eq:fbndup}
    \Jbf(\thetabf) \leq \Ubf\tran\gtilde'(\wbf)\Ubf \leq (1-\epsilon)
    \Ubf\tran\Ubf \leq (1-\epsilon)\Ibf.
\eeq
For the lower bound, observe that
\beqa \label{eq:fbnddown}
    \Jbf(\thetabf) &\stackrel{(a)}{=}& \Ubf\tran\gtilde'(\wbf)\Ubf -
        \left[ \begin{array}{cc} \zero & \zero \\ \zero & \Pbf^\perp \end{array} \right] \nonumber \\
    &\stackrel{(b)}{\geq} & \epsilon \Ubf\tran\Ubf -
        \left[ \begin{array}{cc} \zero & \zero \\ \zero & \Pbf^\perp\end{array} \right] \nonumber \\
    &\stackrel{(c)}{=}&
        \left[ \begin{array}{cc} \epsilon \Pbf & \zero \\ \zero & (\epsilon-1)\Pbf^\perp\end{array} \right] \nonumber \\
    &\stackrel{(d)}{=}&
        \left[ \begin{array}{cc} (\epsilon-1) \Ibf & \zero \\ \zero & (\epsilon-1)\Ibf\end{array} \right]
        +\left[ \begin{array}{cc} \Ibf-\epsilon\Pbf^\perp& \zero \\ \zero & (1-\epsilon)\Pbf\end{array} \right] \nonumber \\
    &\geq& (\epsilon - 1) \Ibf,
\eeqa
where step (a) follows from \eqref{eq:fderiv};
(b) follows from \eqref{eq:gtcontract};
(c) follows from the definition of $\Ubf$ in \eqref{eq:Udef} and the fact that $\Pbf$ and $\Pbf^\perp$
are orthogonal projections;
(d) follows from the definition of $\Pbf^\perp$ in \eqref{eq:Pperp};
and \eqref{eq:fbnddown} follows because the eigenvalues of $\Pbf^\perp$ and $\Pbf$ are in the interval $[0,1]$ and because $\epsilon\in(0,0.5]$.
Together \eqref{eq:fbndup} and \eqref{eq:fbnddown} show that
\[
    \|f'(\thetabf)\|=\|\Jbf(\thetabf)\| \leq 1-\epsilon.
\]
Hence the $f'(\thetabf)$ is a contraction and the
\ALG algorithm converges linearly at rate $1-\epsilon$.

\section{Proof of Theorem \ref{thm:varConv}} \label{sec:varConvPf}

We need to prove that the conditions of Assumption~\ref{as:IterLinConv}
are satisfied.  Property (a) is satisfied since Theorem~\ref{thm:meanConv}
shows that the constrained linearized LSL-BFE optimization
\eqref{eq:FbetheLinOpt} has a unique minima for any
$(\taubf_r,\taubf_p) > 0$.

We next construct the set $\Gamma$.
From the proof of Lemma~\ref{lem:gspConv}, we know that
when $\taubf_x = \var(\xbf|\rbf,\taubf_r)$ and
$\taubf_z = \var(\zbf|\pbf,\taubf_p)$,
\beqa
    && \taubf_x \in \left[
    \frac{A\taubf_r}{A+\taubf_r},\frac{B\taubf_r}{B+\taubf_r} \right],
        \label{eq:tauxInt} \\
    && \taubf_z \in \left[\frac{A\taubf_p}{A+\taubf_p},
    \frac{B\taubf_p}{B+\taubf_p}\right], \label{eq:taurInt}
\eeqa
Hence
\beq \label{eq:tausInt}
    \taubf_s \defn
    \left( 1 - \frac{\taubf_z}{\taubf_p} \right) \frac{1}{\taubf_p}
    \in \left[ \frac{1}{B + \taubf_p}, \frac{1}{A + \taubf_p}
    \right].
\eeq
Now consider a set  $\Gamma$ of the form
\beq \label{eq:tauBnd}
    \Gamma = \left\{(\taubf_r,\taubf_p) ~ \mid ~
    \taubf_r \in [a_r,b_r], \quad \taubf_p \in [a_p,b_p] \right\}.
\eeq
In order that $\Gamma$ satisfies Assumption~\ref{as:IterLinConv}(c),
we need to find bounds $a_r,b_r,a_p,b_p$, such
that if $(\taubf_r,\taubf_p) \in \Gamma$, then
$(\taubfbar_r,\taubfbar_p) \in \Gamma$ where
$(\taubfbar_r,\taubfbar_p)$ are given in \eqref{eq:taubarpr}.

To this end, first observe that \eqref{eq:tauxInt} shows that
$\taubf_x \leq 1/B$, so $\taubfbar_p = \Sbf\taubf_x \leq b_p$
for some $b_p$.
If $\taubf_p \leq b_p$, \eqref{eq:tausInt} shows that
$\taubf_s \in [1/B,1/(A+b_p)]$.
Therefore, using the boundedness assumptions on $\Sbf$,
$\taubfbar_r = \one./(\Sbf\tran \taubf_s) \in [a_r,b_r]$
for some lower and upper bounds $a_r$ and $b_r$.
Finally, if $\taubf_r \geq a_r$, $\taubf_x \geq Aa_r/(A+a_r)$
and hence $\taubfbar_p = \Sbf\taubf_x \geq a_p$ for some $a_p$.
We conclude that we can find bounds
$a_r,b_r,a_p,b_p$, such
that if $(\taubf_r,\taubf_p) \in \Gamma$, then
$(\taubfbar_r,\taubfbar_p) \in \Gamma$, and $\Gamma$
is a compact, convex set satisfying Assumption~\ref{as:IterLinConv}(c).

Finally, we need to show the convexity assumption in
Assumption~\ref{as:IterLinConv}(b).
The linearized LSL-BFE in \eqref{eq:JlinBFE} is separable,
so we only need to consider the convexity
of one of the terms.  To this consider a prototypical term
of the form
\beq \label{eq:Jbscalar}
    J(b) = D(b\|\e^{-f_x}) + \frac{1}{2\tau_r}\var(x|b),
\eeq
where $b(x)$ is some density over a scalar variable $x$ and
$f_x(x)$ is a convex penalty function.
The Hessian of $J(b)$ is a quadratic form that takes
perturbations $v_1(x)$ and $v_2(x)$ to the density $b(x)$
and returns a scalar.  We will denote this Hessian by
$J''(b)(v_1,v_2)$.  Differentiating \eqref{eq:Jbscalar} we obtain
that
\beq \label{eq:Jhess1}
    J''(b)(v,v) = \int \frac{v(x)^2}{b(x)}\d x - \frac{1}{\tau_r}
        \left(\int v(x)x \d x \right)^2.
\eeq
We need to show that this is positive.  For any $v(x)$,
let $u(x) = v(x)/b(x)$ so that $v(x)=u(x)b(x)$.
Since a perturbation
to the density $b(x)$ must satisfy $\int v(x) \d x = 0$, we have
that
\[
    \Exp(u(x)|b) = \int u(x)b(x) \d x = 0.
\]
Also, $J''(b)(v,v)$ above can be written as
\beqa
    J''(b)(v,v) & \stackrel{(a)}{=} &
     \Exp(u(x)^2|b) - \frac{1}{\tau_r}\Exp(u(x)x|b) \nonumber \\
     & \stackrel{(b)}{=} &
     \var(u(x)|b) - \frac{1}{\tau_r}\Exp^2(u(x)(x-\mu_x)|b)  \nonumber \\
     & \stackrel{(c)}{\geq} &
      \var(u(x)|b)\left[1 - \frac{\tau_x}{\tau_r}\right],
\eeqa
where (a) follows from substituting $v(x)=u(x)b(x)$ into \eqref{eq:Jhess1};
in (b) we have used the notation $\mu_x = \Exp(x|b)$
and the fact that $\Exp(u(x)|b) = 0$;
and (c) follows from the Cauchy-Schwartz inequality with
the notation $\tau_x = \var(x|b)$.
Now, using \eqref{eq:tauxInt}, we see that when
$(\taubf_r,\taubf_p) \in \Gamma$, we have the lower bound,
\[
    1 - \frac{\tau_x}{\tau_r} \geq
    1 - \frac{B}{B+a_r} \geq \frac{a_r}{B+a_r} > 0.
\]
We conclude that there exists an $\epsilon$ such that
\[
    J''(b) \geq \epsilon \Ibf,
\]
at any minima $b = \bhat$ to the linearized LSL-BFE when
$(\taubf_r,\taubf_p) \in \Gamma$.
This proves Assumption \ref{as:IterLinConv}(b).
The uniform boundedness of all the other derivatives
follows from the fact that all the terms are twice differentiable
and the set $\Gamma$ is compact.

Thus, all the conditions of Assumption~\ref{as:IterLinConv}
and the theorem follows from Theorem~\ref{thm:iterLinConv}.

\section{Proof of Theorem~\ref{thm:varConvMS} } \label{sec:varConvMSPf}

We begin with proving part (a).  We use induction.  Suppose that
\eqref{eq:initVarConst} is satisfied for some $t$.
Since $\qbf^0$, $\xbf^0$, and $\vbf^0$ are fixed points,
we have from line~\ref{line:qt} of Algorithm~\ref{algo:ALG} that
$\xbf^0=\vbf^0$.
Then, since $\xbf^0$ is a fixed point, we have from lines~\ref{line:rt}
and \ref{line:xzest} and equation \eqref{eq:GxzMS} that
\[
    \xbf^0 = g_x(\rbf^0,\taubf_r^0) = \argmin_{\xbf}
        \fx(\xbf) + \frac{1}{2}\|\xbf-\vbf^0+\taubf_r^{0}.\qbf^0\|_{\taubf_r^{0}}^2.
\]
Therefore, $\xbf=\xbf^0$ is the unique solution to
\[
    \zero = \fx'(\xbf) + \Diag(\one./(2\taubf_r^{0}))(\xbf-\xbf^0) + \qbf^0,
\]
which implies
\beq \label{eq:fxdq0}
    \fx'(\xbf^0) = -\qbf^0.
\eeq
By the induction hypothesis \eqref{eq:initVarConst},
$\xbf^t=\xbf^0$ and $\qbf^t=\qbf^0$.  Since
$\xbf^{\tp1} = g_x(\rbf^t,\taubf_r^t)$, we have $\xbf=\xbf^{\tp1}$
is the unique solution to
\beqa \label{eq:fxdq1}
    \zero
    &=& f'(\xbf) + \Diag(\one./(2\taubf_r^t))(\xbf-\rbf^t) \\
    &=& f'(\xbf) + \Diag(\one./(2\taubf_r^t))(\xbf -\xbf^0) + \qbf^0,
\eeqa
where we have used the fact that
\[
    \rbf^t = \xbf^t + \taubf_r^t.\qbf^t =
    \xbf^0 + \taubf_r^t.\qbf^0.
\]
From \eqref{eq:fxdq0}, $\xbf=\xbf^0$ is also a solution to
\eqref{eq:fxdq1}.
Therefore, $\xbf^{\tp1} = \xbf^0$.  Similarly,
if $\sbf^t=\sbf^0$ and $\zbf^t=\zbf^0$, then $\zbf^{\tp1}=\zbf^0$.
From \eqref{eq:vmin}, $\vbf^{\tp1} = \vbf^0$.
We conclude that if \eqref{eq:initVarConst} is satisfied for some $t$,
it is satisfied for $\tp1$.  So part (a) follows by induction.

To prove part (b), we leverage the convergence result from
\cite{yates:95}.  Using our earlier result \eqref{eq:gxderivMS},
we have that
\[
    \tau_{x_j}^{\tp1} = \tau_{r_j}^t g_{x_j}'(r_j^t,\tau_{r_j}^t) =
    \frac{\tau_{r_j}^t}{1 + \fxj''(x_j^{\tp1})\tau_{r_j}^t}.
\]
Rewriting this in vector form and using the updates in Algorithm~\ref{algo:ALG} with $\theta^t = 1$, we obtain that
\beqa
   \one./\taubf_x^{\tp1}
   &=& \one./\taubf_r^t + \fx''(\xbf^{\tp1})
   = \Sbf\tran\taubf_s^t + \fx''(\xbf^{\tp1}) \nonumber\\
   &=& \Sbf\tran\taubf_s^t + \xibf_x, \quad
   \xibf_x \defn \fx''(\xbf^{\tp1}) > \zero \label{eq:tauxit0}
\eeqa
\textb{where $\fx''(\xbf) = [f_{x_1}''(x_1),\dots,f_{x_n}''(x_n)]\tran$ and}
where $\xibf_x$ is positive due to the convexity
assumption and invariant to $t$ due to part (a).
Similarly, for the output estimation function $g_z$,
\[
    \taubf_z^{\tp1} = \taubf_p^t . g_z'(\pbf^t,\taubf_p^t)
    = \taubf_p^t./(\one + \fz''(\zbf^{\tp1}) . \taubf_p^t).
\]
Therefore, from the modified update of $\taubf_s^{\tp1}$ in
\eqref{eq:taustMS},
\[
    \taubf_s^{\tp1}
    = \fz''(\zbf^{\tp1}) ./(
    \one + \fz''(\zbf^{\tp1}) . \taubf_p^t),
\]
or equivalently,
\beqa \label{eq:tausit0}
    \one./\taubf_s^{\tp1}
    &=&\taubf_p^t + \one./\fz''(\zbf^{\tp1}) = \Sbf\taubf_x^{t} + \xibf_z \\
    \xibf_z &\defn& \one./\fz''(\zbf^{\tp1}).
\eeqa

Now define the maps,
\beqan
    \Phi_s(\taubf_x) &:=& \one./\big( \Sbf\taubf_x + \xibf_z\big) \\
    \Phi_x(\taubf_s) &:=& \one./\big( \Sbf\tran\taubf_s + \xibf_x \big)
\eeqan
so that the updates \eqref{eq:tausit0} and \eqref{eq:tauxit0}
can be written as
\[
    \taubf_s^{t} = \Phi_s(\taubf_x^{\tm1}), \quad
    \taubf_x^{\tp1} = \Phi_x(\taubf^t_s).
\]
Note that, due to part (a), $\xibf_x$ and $\xibf_z$ in \eqref{eq:tauxit0}
and \eqref{eq:tausit0}, do not change with $t$.
It is easy to check that, for any $\Sbf > 0$,
\begin{enumerate}
\item[(i)] $\Phi_s(\taubf_x) > 0$,
\item[(ii)] $\taubf_x \geq \taubf_x' \Rightarrow
    \Phi_s(\taubf_x) \leq     \Phi_s(\taubf_x')$, and
\item[(iii)] For all $\alpha > 1$,
    $\Phi_s(\alpha\taubf_x) > (1/\alpha)\Phi_s(\taubf_x)$.
\end{enumerate}
with the analogous properties being satisfied by $\Phi_x(\taubf_s)$.
Now let $\Phi := \Phi_x\circ \Phi_s$ be the composition
of the two functions so that $\taubf^{\tp1}_x = \Phi(\taubf_x^{\tm1})$.
Then, $\Phi$ satisfies the three properties:
\begin{enumerate}
\item[(i)] $\Phi(\taubf_x) > 0$,
\item[(ii)] $\taubf_x \geq \taubf_x' \Rightarrow
    \Phi(\taubf_x) \geq  \Phi(\taubf_x')$, and
\item[(iii)] For all $\alpha > 1$,
    $\Phi(\alpha\taubf_x) < \alpha\Phi(\taubf_x)$.
\end{enumerate}
Also, for any $\taubf_s\geq 0$, we have $\Phi_x(\taubf_s) \leq \one./\xibf_x$,
and therefore, $\Phi(\taubf_x) \leq \one./\xibf_x$ for all $\taubf_x\geq 0$.
Hence, taking any $\taubf_x \geq \one./\xibf_x$, we obtain:
\[
    \taubf_x \geq \Phi(\taubf_x).
\]
The results in \cite{yates:95} then show that the updates
$\taubf^{\tp1}_x = \Phi(\taubf_x^{\tm1})$ converge to unique fixed points.
Since the increment increases by two, we need to apply
the convergence twice:  once for the $\taubf^t_x$ with odd values
of $t$, and a second time for even values.  Since the limit points are
unique, both the even and odd sub-sequences will converge to the same value.
A similar argument shows that $\taubf_s^t$ also converges
to unique fixed points.

\section{Original GAMP via Stale, Linearized ADMM} \label{sec:original}

First, we examine the minimization in \eqref{eq:LADMMz}.
Starting with \eqref{eq:LaugGAMP}, a derivation identical to \eqref{eq:saug},
but with $\xbf^{\tp1}=\Exp(\xbf|b_x^{\tp1})$ in place of $\vbf$, yields
\beqa
    \lefteqn{ L(b_x^{\tp1},b_z,\sbf^t;\taubf_p) }\nonumber\\
    &=& J(b_x^{\tp1},b_z,\taubf_r,\taubf_p) + (\sbf^t)\tran(\Exp(\zbf|b_z)-\Abf\xbf^{\tp1})
        \nonumber \\ && \mbox{}
    	+ \tfrac{1}{2}\big\|\Exp(\zbf|b_z)-\Abf\xbf^{\tp1}\big\|^2_{\taubf_p} \\
    &=& D(b_z\| \Zz^{-1}\e^{-\fz})
        + \left(\one./(2\taubf_p)\right)\tran \var(\zbf|b_z)
        \label{eq:LaugGAMPz} \\ && \mbox{}
    	+ \Exp \big( \tfrac{1}{2}
    	\|\zbf-(\Abf\xbf^{\tp1}-\taubf_p.\sbf^t)\|_{\taubf_p}^2 \big| b_z \big)
	- \sum_{i=1}^m \frac{ \tau_{z_i} }{2\tau_{p_i}}
    	+ \const , \nonumber \\
    &=& D(b_z\| \Zz^{-1}\e^{-\fz})
    	+ \tfrac{1}{2}
    	\Exp \big( \|\zbf-(\Abf\xbf^{\tp1}-\taubf_p.\sbf^t)\|_{\taubf_p}^2 \big| b_z \big)
        \nonumber \\ && \mbox{}
    	+ \const , \nonumber \\
    &=& \int_{\R^m} b_z(\zbf) \ln \tfrac{b_z(\zbf)}{
        \exp(
	-\fz(\zbf) - \frac{1}{2} \|\zbf-(\Abf\xbf^{\tp1}-\taubf_p.\sbf^t)\|_{\taubf_p}^2
	) } \d \zbf
	\nonumber\\&&\mbox{}
    	+ \const \\
    &=& D(b_z\|p_z) + \const ,
\eeqa
where ``\const" is constant with respect to $b_z$ and
$p_z(\zbf)\propto \exp(-\fz(\zbf) - \frac{1}{2} \|\zbf-(\Abf\xbf^{\tp1}-\taubf_p.\sbf^t)\|_{\taubf_p}^2)$.
Thus, the minimizing density $b_z$ output by \eqref{eq:LADMMz} is
\beqa
    b_z^{\tp1}(\zbf)
    &\propto& \exp\big( -\fz(\zbf)
	- \tfrac{1}{2} \|\zbf-\pbf^{\tp1}\|_{\taubf_p}^2 \big) \\
    \pbf^{\tp1}
    &\defn& \Abf\xbf^{\tp1}-\taubf_p.\sbf^t .
    \label{eq:p}
\eeqa

Next we examine the minimization in \eqref{eq:LADMMx}.
The objective function in \eqref{eq:LADMMx} can be written, using
\eqref{eq:LaugGAMP}, $\xbf^{t}=\Exp(\xbf|b_x^{t})$, and \eqref{eq:JlinBFE},
as follows:
\beqa
    \lefteqn{ L(b_x,b_z^t,\sbf^{\tm1};\taubf_p) }\nonumber\\
    \lefteqn{
    	+ \tfrac{1}{2}\big(\Exp(\xbf|b_x)-\xbf^t\big)\tran
	\big(\Dbf_{\taubf_r}-\Abf\tran\Dbf_{\taubf_p}\Abf\big)
	\big(\Exp(\xbf|b_x)-\xbf^t\big)
	}\nonumber\\
    &=& D(b_x\| \e^{-\fx})
        + \big(\one./\taubf_r\big)\tran \var(\xbf|b_x)
    	- (\sbf^{\tm1})\tran \Abf \Exp(\xbf|b_x)
        \nonumber \\ && \mbox{}
        + \tfrac{1}{2}\Exp(\xbf|b_x)\tran\Abf\tran
		\Dbf_{\taubf_p}\Abf\Exp(\xbf|b_x)
	- (\zbf^t)\tran \Dbf_{\taubf_p}\Abf\Exp(\xbf|b_x)
        \nonumber \\ && \mbox{}
    	+ \tfrac{1}{2}\Exp(\xbf|b_x)\tran
	\big(\Dbf_{\taubf_r}-\Abf\tran\Dbf_{\taubf_p}\Abf\big)
	\Exp(\xbf|b_x)
        \nonumber \\ && \mbox{}
    	-(\xbf^t)\tran
	\big(\Dbf_{\taubf_r}-\Abf\tran\Dbf_{\taubf_p}\Abf\big)
	\Exp(\xbf|b_x)
    	+ \const \nonumber \\
    &=& D(b_x\| \e^{-\fx})
        + \big(\one./\taubf_r\big)\tran \var(\xbf|b_x)
    	- (\sbf^{\tm1})\tran \Abf \Exp(\xbf|b_x)
        \nonumber \\ && \mbox{}
	- (\zbf^t)\tran \Dbf_{\taubf_p}\Abf\Exp(\xbf|b_x)
    	+ \tfrac{1}{2}\Exp(\xbf|b_x)\tran \Dbf_{\taubf_r} \Exp(\xbf|b_x)
        \nonumber \\ && \mbox{}
    	-(\xbf^t)\tran \Dbf_{\taubf_r} \Exp(\xbf|b_x)
    	+(\xbf^t)\tran \Abf\tran\Dbf_{\taubf_p}\Abf \Exp(\xbf|b_x)
    	+ \const \nonumber \\
    &\stackrel{(a)}{=}& D(b_x\| \e^{-\fx})
        + \big(\one./\taubf_r\big)\tran \var(\xbf|b_x)
    	- (\sbf^t)\tran \Abf \Exp(\xbf|b_x)
        \nonumber \\ && \mbox{}
    	+ \tfrac{1}{2}\Exp(\xbf|b_x)\tran \Dbf_{\taubf_r} \Exp(\xbf|b_x)
    	-(\xbf^t)\tran \Dbf_{\taubf_r} \Exp(\xbf|b_x)
    	+ \const \nonumber \\
    &\stackrel{(b)}{=}& D(b_x\| \e^{-\fx})
        + \big(\one./\taubf_r\big)\tran \var(\xbf|b_x)
        \nonumber \\ && \mbox{}
    	+ \tfrac{1}{2}\Exp(\xbf|b_x)\tran \Dbf_{\taubf_r} \Exp(\xbf|b_x)
    	-(\rbf^t)\tran \Dbf_{\taubf_r} \Exp(\xbf|b_x)
    	+ \const \nonumber \\
    &=& D(b_x\| \e^{-\fx})
        + \big(\one./\taubf_r\big)\tran \var(\xbf|b_x)
    	+ \tfrac{1}{2}\|\Exp(\xbf|b_x)-\rbf^t \|_{\taubf_r}^2
        \nonumber \\ && \mbox{}
    	+ \const \nonumber \\
    &=& D(b_x\| \e^{-\fx})
    	+ \Exp\big(\tfrac{1}{2}\|\xbf-\rbf^t \|_{\taubf_r}^2\big|b_x\big)
        + \const ,
	\label{eq:LaugGAMPx} \\
    &=& \int_{\R^n} \!\!\! b_x(\xbf) \ln \tfrac{b_x(\xbf)}{
        \exp(
	-\fx(\xbf) - \frac{1}{2} \|\xbf-\rbf^t\|_{\taubf_r}^2
	) } \d\xbf
    	+ \const \\
    &\stackrel{(c)}{=}& D(b_x\|p_x) + \const,
\eeqa
where ``\const" is constant with respect to $b_x$;
line (a) used \eqref{eq:LADMMs};
line (b) used
\beqa
    \rbf^t
    &\defn& \xbf^t + \Diag(\taubf_r)\Abf\tran\sbf^t ;
\eeqa
and line (c) used
$p_x(\xbf)\propto \exp(-\fx(\xbf) - \frac{1}{2} \|\xbf-\rbf^t\|_{\taubf_p}^2)$.
Thus, the minimizing density $b_x$ output by \eqref{eq:LADMMx} is
\beqa
    b_x^{\tp1}(\xbf)
    &\propto& \exp\big( -\fx(\xbf)
	- \tfrac{1}{2} \|\xbf-\rbf^t\|_{\taubf_r}^2 \big) .
\eeqa

Finally, using \eqref{eq:p} in \eqref{eq:LADMMs}, we obtain
\beq
    \sbf^{\tp1}
    = (\zbf^{\tp1}-\pbf^{\tp1})./\taubf_p .
\eeq
Thus, we have recovered the mean updates of the original sum-product
GAMP algorithm, i.e., the non-indented lines in Algorithm~\ref{algo:GAMP}.

\bibliographystyle{IEEEtran}
\bibliography{bibl}
\end{document}